\RequirePackage{silence}
\documentclass[fleqn,usenatbib,useAMS]{mnras} 
\usepackage{graphicx}
\usepackage{amsmath}	
\usepackage{amssymb}

\usepackage{tikz}
\usepackage{braket}
\usepackage{url}
\usepackage{array}
\usepackage{multirow}

\usepackage{amsfonts}
\usepackage{float}
\usepackage{bm}
\usepackage{cancel}
\usepackage{ae,aecompl}
\usepackage{array}
\usepackage{soul}
\usepackage{mathtools}
\usepackage{multirow}
\usepackage[utf8]{inputenc}
\usepackage{booktabs}
\usepackage{graphicx}
\usepackage{substitutefont}
\substitutefont{TS1}{aer}{cmr}
\usepackage{makecell}

\newtheorem{definition}{Definition}[section]
\DeclareMathAlphabet{\pazocal}{OMS}{zplm}{m}{n}

\usepackage{calrsfs} 

\DeclareRobustCommand{\appropto}{\mathrel{\vcenter{
		\offinterlineskip\halign{\hfil$##$\cr 
			\propto\cr\noalign{\kern2pt}\sim\cr\noalign{\kern-2pt}}}}}

\title[Distinct RARs of galaxies and clusters in HMG]{Distinct radial acceleration relations of galaxies and galaxy clusters supports hyperconical modified gravity} 

\author[R. Monjo \& I. Banik]{Robert Monjo$^1$\thanks{Email: \href{mailto:rmonjo@ucm.es}{rmonjo@ucm.es} (Robert Monjo)} \vspace{10pt} and Indranil Banik$^{2}$\thanks{Email: \href{mailto:indranilbanik1992@gmail.com}{indranilbanik1992@gmail.com} (Indranil Banik)}  \\
$^{1}$Department of Algebra, Geometry and Topology, Complutense University of Madrid, Pza. Ciencias 3, E-28040 Madrid, Spain.
\\
$^{2}$Scottish Universities Physics Alliance, University of Saint Andrews, North Haugh, Saint Andrews, Fife, KY16 9SS, UK.
}

\pubyear{2024}
\begin{document}
\label{firstpage}

\pagerange{\pageref{firstpage}--\pageref{lastpage}}

\maketitle

\begin{abstract}
General relativity (GR) is the most successful theory of gravity, with great observational support on local scales. However, to keep GR valid over cosmic scales, some phenomena (such as flat galaxy rotation curves and the cosmic expansion history) require the assumption of exotic dark matter. The radial acceleration relation (RAR) indicates a tight correlation between dynamical mass and baryonic mass in galaxies. This suggests that the observations could be better explained by modified gravity theories without exotic matter. Modified Newtonian Dynamics (MOND) is an alternative theory that was originally designed to explain flat galaxy rotation curves by using a new fundamental constant acceleration $a_0$, the so-called Milgromian parameter. However, this non-relativistic model is too rigid (with insufficient parameters) to fit the large diversity of observational phenomena. In contrast, a relativistic MOND-like gravity naturally emerges from the hyperconical model, which derives a fictitious acceleration compatible with observations. This study analyses the compatibility of the hyperconical model with respect to distinct RAR observations of 10 galaxy clusters obtained from HIFLUGCS and 60 high-quality SPARC galaxy rotation curves. The results show that a general relation can be fitted to most cases with only one or two parameters, with an acceptable $\chi^2$ and $p$-value. These findings suggest a possible way to complete the proposed modification of GR on cosmic scales.
\end{abstract}

\begin{keywords}
    galaxies: clusters: general -- gravitation -- methods: statistical -- galaxies: kinematics and dynamics -- methods: analytical
\end{keywords}

\section{Introduction}
\label{sec:intro}

\subsection{The missing gravity problem}

As is well known, observational tests of General Relativity (GR) show successful results on Solar System scales \citep{Dittus2007, Ciufolini2019, Touboul2022, Liu2022, Desmond2024, Vokrouhlicky_2024}. The success of standard gravity seems to be in question only on larger scales \citep{Chae2020, Banik2022}. It is well-known that exotic cold dark matter (CDM) is required to extend GR to cosmic scales. However, the hypothetical CDM particles present strong challenges, in particular the tight empirical relationship between observed gravitational anomalies and the distribution of visible baryonic matter in galaxies \citep{Trippe2014, Merritt2017, Goddy2023}. This empirical law is known as the mass-discrepancy acceleration relation \citep[MDAR;][]{McGaugh2004, DiCintio2015, Desmond2016}, the mass-luminosity relation \citep{Leauthaud2010, Cattaneo2014}, the baryonic Tully-Fisher relation \citep[BTFR;][]{Lelli2019, Goddy2023}, or the more general radial acceleration relation \citep[RAR;][]{McGaugh2016, Lelli2017, Tian2020}.

\citet{Tian2020} found that the observed RAR in galaxy clusters is consistent with predictions from a semi-analytical model developed in the standard Lambda-CDM \citep[$\Lambda$CDM;][]{Efstathiou1990, Ostriker1995} framework. To explain how the contribution of CDM is determined by that of baryons, some authors suggest that they present a strong coupling that leads to an effective law such as the MDAR/BTFR/RAR \citep{Blanchet2007, Katz2016, Barkana2018}. It has also been argued that such a correlation can arise in the $\Lambda$CDM framework once baryonic feedback effects are simulated with adequate resolution \citep{Mercado2024}.

The lack of direct or indirect non-gravitational detection of dark matter suggests a weak or even non-existent coupling between CDM and baryons \citep{Abel2017, Hoof2020, Du2022, Aalbers2023, Hu2024}, which is in conflict with these empirical relationships. Moreover, excess rotation occurs only where the Newtonian acceleration $a_{N}$ induced by the visible matter satisfies $a_{N} \la a_0 \approx 1.2 \times 10^{-10} \mathrm{m\,s}^{-2}$, suggesting that the missing gravity problem is a space-time problem rather than a matter-type problem. This is also consistent with the deficient dark-matter halos that some relic galaxies seem to have despite being above the $a_0$ scale \citep{Comeron2023}. In other cases, the CDM halo hypothesis also predicts a systematically deviating relation from the observations, with densities about half of what is predicted by CDM simulations \citep{deBlok2008}. In general, galaxy rotation curves and velocity dispersions appear to be more naturally explained by modified gravity \citep{McGaugh2007, McGaugh2016, Famaey2012, Banik2022, Chae2022}. 

The dark matter hypothesis also presents difficulties in explaining some phenomena such as the absence of the expected \textit{Chandrasekhar dynamical friction} in cluster collisions, falsified by more than $7\sigma$ \citep{Kroupa2015,Ardi2020,Kroupa2023}. The lack of dynamical friction on galaxy bars is a strong argument that the central density of CDM in typical disc galaxies has to be a lot smaller than expected in standard CDM simulations \citep{Roshan2021}. Another example is the morphology of dwarf galaxies. According to \citet{Asencio2022}, theobserved deformations of dwarf galaxies in the Fornax Cluster and the lack of low surface brightness dwarfs near its centre are incompatible with $\Lambda$CDM predictions. Moreover, the dwarfs analyzed in that study have sufficiently little stellar mass that the observations cannot be explained by baryonic feedback effects, but they are consistent with the Milgromian modified Newtonian dynamics \citep[MOND;][]{Milgrom1983}. Therefore, most observations suggest the need to explore modified gravity as an alternative to the standard model \citep{Trippe2014, Merritt2017}.

\subsection{Beyond the MOND paradigm}

The MOND paradigm has been deeply explored from galactic dynamics to the \textit{Hubble tension}, which is explained by a more efficient (early) formation of large structures such as the local supervoid \citep{Keenan2013,Haslbauer2020,Banik2022,Mazurenko2023}. The RAR has been thoroughly analyzed for galaxy rotation curves collected from the Spitzer Photometry and Accurate Rotation Curves (SPARC) sample \citep{Lelli2016, Lelli2019}. The results were anticipated over three decades ago by MOND \citep{Milgrom1983, McGaugh2016}, although the form of the transition between the Newtonian and Milgromian regimes must be found empirically.


However, the relativistic formulation of MOND has been less successful. In particular, Bekenstein proposed a non-cosmological version of Tensor-Vector-Scalar (TeVeS) gravity \citep{Bekenstein2004, Famaey2012} that predicts unstable stars on a timescale of a few weeks \citep{Seifert2007}, which is only avoidable with an undetermined number of terms \citep{Mavromatos2009}. To solve these issues, \citet{Skordis2021} found that, by adding terms analogous to the FLRW action, at least the second-order expansion is free of ghost instabilities. Their model is also capable of obtaining gravitational waves traveling at the speed of light $c$, which was not the case with the original TeVeS. However, the authors pointed out that it needs to be embedded in a more fundamental theory. 

Recently, \citet{Blanchet2024} proposed a relativistic MOND formulation based on space-time foliation by three-dimensional space-like hypersurfaces labeled by the Khronon scalar field. The idea is very similar to the Arnowitt–Deser–Misner (ADM) treatment in the dynamical embedding of the hyperconical universe \citep{Monjo2017, Monjo2018, MCS2020, Monjo2023, Monjo2024, Monjo2024b}. 

Applying perturbation theory to the hyperconical metric, a relativistic theory with MOND phenomenology is obtained, which adequately fits 123 SPARC galaxy rotation curves \citep{Monjo2023}. The cosmic acceleration derived from it is $a_{\gamma 0} \equiv 2\gamma_0^{-1}c/t$, where $t$ is the age of the universe and $\gamma_0>1$ is a projection parameter that translates from the ambient spacetime to the embedded manifold \citep{MCS2023, Monjo2024}. In contrast to the Milgrom constant $a_0$, the cosmic acceleration $a_{\gamma 0}$ is a variable that depends on the geometry considered, especially the ratio between the Keper-Newton orbital speed and the Hubble flux. Numerical equivalence between the $a_0$ and $a_{\gamma 0}$ scales is found for $\gamma_0 \approx 13 \pm 3$ or equivalently for $\gamma_0^{-1} \approx 0.08 \pm 0.02$. 

In the limit of weak gravitational fields and low velocities, the hyperconical model is also linked to the scalar tensor vector gravity (STVG) theory, popularly known as Moffat gravity \citep[MOG;][]{Moffat2006}. The MOG/STVG model is a fully covariant or Lorentz invariant theory that includes a dynamical massive vector field and scalar fields to modify GR with a dynamical `gravitational constant' $G$ \citep{Moffat2009, Moffat2013, Harikumar2022}. In particular, MOG leads to an anomalous acceleration of about $2\, G \alpha_G D^2  \, \approx \,  1.1 \times 10^{-10} \mathrm{m\,s}^{-2}  \, \approx \, 2\gamma_0^{-1}c/t$ for $\gamma_0 \approx 12$, with $\alpha_G \approx 10$ and the universal MOG constant $D = 6.25 \times 10^3 M_{\odot}^{1/2}\;\mathrm{kpc}^{-1}$. Fixing these parameters using galaxy rotation curves, MOG fails to account for the observed velocity dispersion profile of Dragonfly 44 at $5.5\sigma$ confidence, even if one allows plausible variations to its star formation history and thus stellar mass-to-light ratio \citep{Haghi2019DF44}. MOG also struggles to explain the galactic rotation curve, where the discrepancy is smaller, but the measurements are more accurate \citep{Negrelli2018}.

The number of parameters needed to accommodate most theories to the observations of galaxy clusters is perhaps too large and unnatural. In all cases, the phenomenological parameters (e.g., the CDM distribution profile, the ad-hoc MOND interpolating function $\mu$, and the MOG constant $D$) need additional theoretical motivation. In contrast, the hyperconical model proposed by Monjo derives a natural modification to GR from minimal dynamical embedding in a (flat) five-dimensional Minkowskian spacetime \citep{Monjo2023, Monjo2024, Monjo2024b}.

Therefore, this paper aims to show how the anomalous RAR in ten galaxy clusters analysed by \citet{Eckert2022} and \citet{Li2023} can be adequately modeled by the hyperconical modified gravity (HMG) of \citet{Monjo2023}. As \citet{Tian2020} pointed out, clusters present a larger anomalous acceleration ($g \sim 10^{-9} \mathrm{m\,s}^{-2}$) than galaxy rotation curves ($g \sim 10^{-10} \mathrm{m\,s}^{-2}$), reflecting the missing baryon problem that remains a challenge for MOND in galaxy clusters \citep{Famaey2012, Li2023, Tian2024}. This open issue is addressed here with the following structure: Section~\ref{sec:obs_and_model} summarizes the data used and the HMG model; Section~\ref{sec:results} shows the main results in the fits and discusses predictions for galaxies and smaller systems, and finally Section~\ref{sec:conclusions} points out the most important findings and concluding remarks.



\section{Data and model}
\label{sec:obs_and_model}
\subsection{Observations used}

We use observational estimates of the radial acceleration relations (RAR; total gravity observed compared to Newtonian gravity due to baryons) for 10 galaxy clusters that were collected from the HIghest X-ray FLUx Galaxy Cluster Sample \citep[HIFLUGCS;][]{Li2023}. In particular, the galaxy clusters considered are as follows: A0085, A1795, A2029, A2142, A3158, A0262, A2589, A3571, A0576, A0496. These clusters have redshift $z$ in the range $0.0328 - 0.0899$. We also compare our results to rotation curves collected from 60 high-quality SPARC galaxies filtered to well-measured intermediate radii \citep{McGaugh2007, McGaugh2016, Lelli2019}.

\subsection{RAR from hyperconical modified gravity (HMG)}

Our main aim is to assess whether the empirical RAR agrees with HMG as developed by \citet{Monjo2023} and summarised here. Let $g$ be the background metric of the so-called hyperconical universe \citep{Monjo2017, Monjo2018, MCS2020, MCS2023}. Working in units where $c = 1$ and the age of the universe is $t_0 = 1$, the metric $g$ is locally approximately given by
\begin{equation} 
g \approx dt^2 (1- kr'^2 ) 
-  \frac{t^2}{t_{0}^2} \left( \frac{dr'^2}{1-kr'^2} + {r'}^2d{\Sigma}^2 \right)
-  \frac{2r't dr'dt}{t_{0}^2\sqrt{1-kr'^2}}, \label{eq:hyp1} 
\end{equation}
where $k = 1/t_0^2$ is the spatial curvature for the current age $t$ of the universe, $t/{t_0}$ is the scale factor associated with lengths because the expansion history is linear \citep{Monjo2024}, $r' \ll t_0$ is the comoving distance, and $\Sigma$ represents the angular coordinates. The shift and lapse terms in Equation~\ref{eq:hyp1} lead to an apparent radial spatial inhomogeneity that appears as a fictitious acceleration with adequate stereographic projection coordinates, which is a candidate to explain the Hubble tension \citep{MCS2023}.

Any gravitational system of mass $M_{sys}$ generates a perturbation over the background metric (Equation~\ref{eq:hyp1}) that can be written as $g \to \hat{g}$ such that $k{r'^2} \to \hat{k}{{\hat r}'^2} \equiv k{{\hat r}'^2} +{2GM_{sys}}/{{\hat r}'}$. Applying local validity of GR (Appendix~\ref{annex:A}), the perturbation term $\hat{h} \equiv \hat{g}-g$ is a key aspect of the HMG model (Appendix~\ref{sec:hyperconical}). Also important is the stereographic projection of the coordinates $r' \to {\hat r}' = \lambda^{1/2}r'$ and $t \to \hat t = \lambda t$, both of which are given by a scaling factor $\lambda \equiv 1/(1-\gamma/\gamma_0)$ that is a function of the angular position $\gamma = \sin^{-1} \left( r'/t_0 \right)$ and a projection factor $\gamma_0^{-1} = \gamma_{sys}^{-1}\cos \gamma_{sys}$, where $\gamma_{sys}$ is the characteristic angle of the gravitational system (Appendix~\ref{sec:gravitational_angle}). In an empty universe, $\gamma_0 = \gamma_U / \cos \gamma_U$. We expect $\gamma_U = \mathrm{\pi}/3$ and therefore $\gamma_0 = 2\mathrm{\pi}/3 \approx 2$. The projection factor of maximum causality, $\gamma_0^{-1} = 1$, arises for $\gamma_U \approx 0.235\mathrm{\pi}$ as then $\gamma_U = \cos \gamma_U$.

When geodesic equations are applied to the projected time component of the perturbation $\hat{h}_{tt}$, a fictitious cosmic acceleration of roughly $\gamma_0^{-1}c/t$ emerges in the spatial direction (see Appendix~\ref{sec:fist_pert_geod}):
\begin{eqnarray}
    \frac{\left| a_{tot} - a_N \right|}{c/t} \approx  \frac{1}{\gamma_0}  \approx  \frac{\cos\gamma_{sys}}{\gamma_{sys}} \, ,
    \label{eq:RAR}
\end{eqnarray}
where $a_N \equiv GM_{sys}/r^2$ is the Newtonian acceleration. However, a time-like component is also found in the acceleration that contributes to the total centrifugal acceleration $a_C$ such that (see Equation~\ref{eq:RAR_aNc} of Appendix~\ref{sec:fist_pert_geod})
\begin{eqnarray}
    a_{C} \approx  \sqrt{a_N^2 + |a_N|\frac{2c}{\gamma_0 t}} \, ,
\end{eqnarray}
which is useful to model galaxy rotation curves under the HMG framework \citep{Monjo2023}. Alternatively to Equation~\ref{eq:RAR}, the cluster RAR is usually expressed as a quotient between total and Newtonian (spatial) acceleration.
\begin{eqnarray}
    \frac{a_{tot}}{a_N} \approx 1 + \frac{c}{a_N \gamma_0 t} \, ,
    \label{eq:RAR_aN}
\end{eqnarray}
with factor $\gamma_0^{-1} = \gamma_{sys}^{-1}\cos \gamma_{sys}$, where the projective angle $\gamma_{sys}$ can be estimated from the cluster approach (Equation~\ref{eq:model_clustb}) or from the general model (Equation~\ref{eq:model_generalb}), respectively, by considering the relative geometry (angle) between the Hubble speed $v_H \equiv r/t$ and the Newtonian circular speed $v_{N} \equiv \sqrt{GM_{sys}/r}$. This can be done as follows:
\begin{eqnarray}
    \label{eq:model_clust2}
    & \sin^2 \gamma_{sys}(r) \approx \nonumber \\
    & \approx \sin^2\gamma_{cen} - \left( \sin^2\gamma_{cen} - \sin^2 \gamma_U \right) \frac{2v_N^2\left( r \right)}{\varepsilon^2_H v_H^2\left( r \right) + 2v_N^2\left( r \right)} \nonumber \\
\end{eqnarray}
\begin{eqnarray}
     & \sin^2 \gamma_{sys}(r) \approx \nonumber 
     \\
     &\approx \sin^2\gamma_U + \left( \sin^2\gamma_{cen} - \sin^2 \gamma_U \right) \left| \frac{2v_N^2 \left( r \right) - \epsilon^2_H v_H^2\left( r \right)}{2v_N^2\left( r \right) + \epsilon^2_H v_H^2\left( r \right)} \right|,
    \label{eq:model_general2}
\end{eqnarray}
where $\gamma_{sys} \left( r \right)$ is the projective angle of the gravitational system (galaxy or cluster), $\gamma_{cen}$ is the gravitational angle of its central element (black hole or brightest galaxy, respectively), and the parameter 
$\epsilon^2_H$ is the so-called \textit{relative density of the neighborhood} (Appendix~\ref{sec:projective_angle}). $\gamma_{cen} \approx \mathrm{\pi}/2$ and $\gamma_U = \mathrm{\pi}/3$ or $\gamma_U \approx 0.235\mathrm{\pi}$ can be fixed here to set a 1-parameter ($\epsilon_H$) general model from Equation~\ref{eq:model_general2}. As a second-order approach, this study also assumes that $\left(\epsilon_H, \gamma_{cen}\right)$ can be free in our 2-parameter model for clusters (Equation~\ref{eq:model_clust2}).

\section{Results and discussion}
\label{sec:results}

\subsection{Fitted values}

Individually, fitting of Equation~\ref{eq:model_clust2} for the quotient between total and Newtonian acceleration (Equation~\ref{eq:RAR_aN}) leads to a square root of the relative density of about $\varepsilon_H = 38^{+29}_{-11}$ (90\% confidence level; Appendix~\ref{sec:individual_fitting} and Figure~\ref{fig:FigC2}). Using the specific model for clusters (Equation~\ref{eq:model_clust2}), all fits provide an acceptable $\chi^2$ ($p$-value $< 0.667$) except for cluster A2029, which did not pass the $\chi^2$ test for the fixed \textit{neighborhood projective angle} of $\gamma_0 = 2$ (i.e., $\gamma_U = \mathrm{\pi}/3$). However, it did for $\gamma_0 = 1$ (i.e., $\gamma_U \approx 0.235\mathrm{\pi}$), which implies we need to use the causality limit for the cosmic acceleration instead of the empty-space limit.

Globally, the correlation of RAR values (differences) with respect to the Newton-Hubble speed ratio approach (Equations~\ref{eq:RAR} and \ref{eq:model_clust2}) is slightly higher ($R^2 = 0.83$) than with respect to using the Newtonian acceleration ($R^2 = 0.79$). The simplest model of fixing $\gamma_U = \mathrm{\pi}/3$ and using a single global parameter of $\varepsilon_H = 40_{-6}^{+8}$ gives a Pearson coefficient of $R^2 = 0.75$, while if $\gamma_U = \mathrm{\pi}/3$ is replaced by $\gamma_U = 0.235\mathrm{\pi}$, we get instead that $R^2 = 0.83$ with $\varepsilon_H = 60_{-8}^{+20}$ (90\% confidence level). 

\begin{figure*}
    \includegraphics[width=\textwidth]{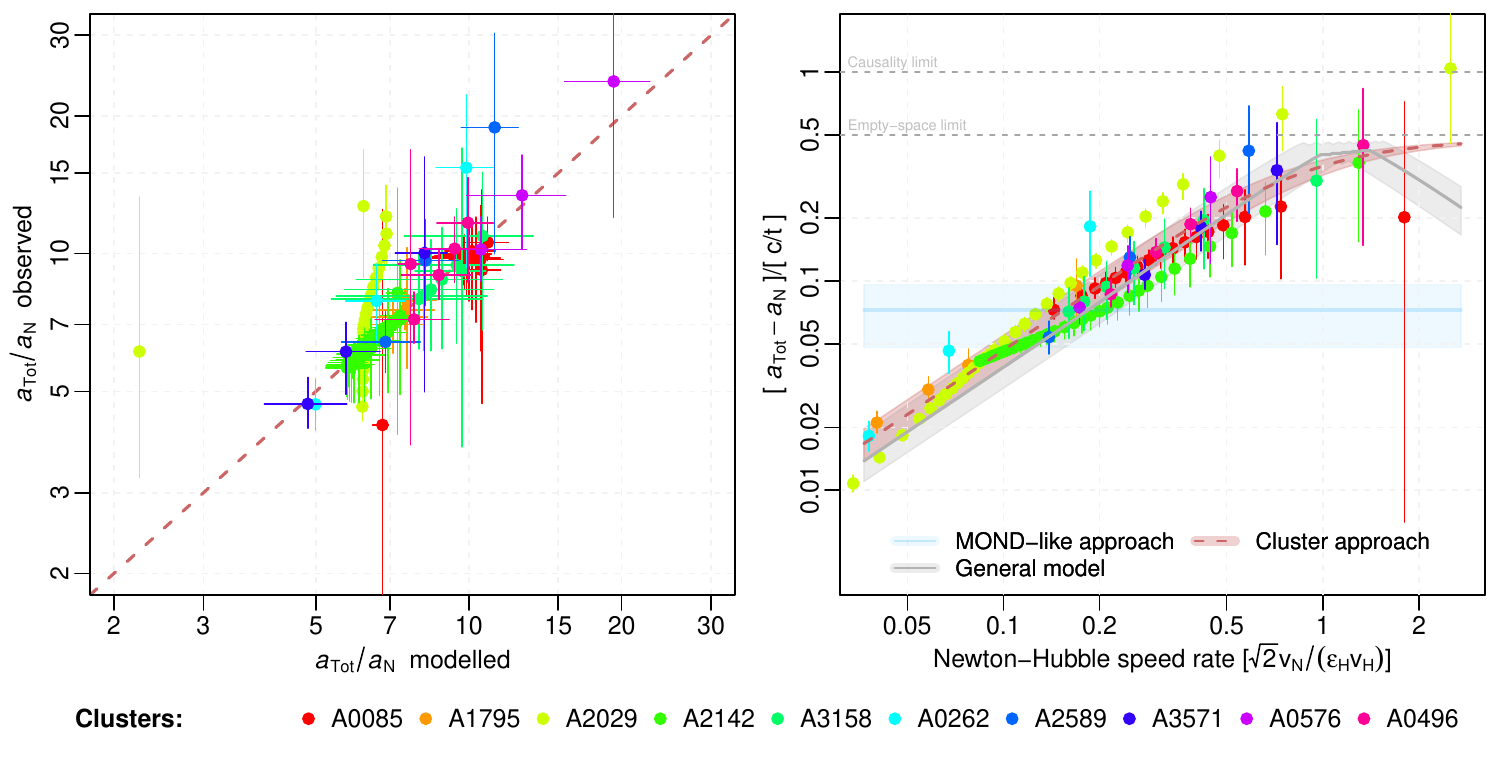}
    \caption{RAR modelling with HMG for the total acceleration ($a_{tot}$) compared to the Newtonian acceleration ($a_{N}$). \emph{Left}: Individual fitting (Equations~\ref{eq:RAR_aN} and \ref{eq:model_clust2}) for the galaxy clusters considered \citep{Li2023}, using two parameters ($\varepsilon_H$ and $\gamma_{cen}$; Table~\ref{table:table2}). \emph{Right}: Global fitting for all data according to three models: MOND-like constant (blue band), general model (grey band, Equations~\ref{eq:RAR} and \ref{eq:model_general2}), and the cluster model (red band, Equations~\ref{eq:RAR} and \ref{eq:model_clust2}). The MOND-like model with constant $\gamma_0^{-1} = \gamma_{sys}^{-1}\cos\gamma_{sys}$ was considered with $\gamma_{sys} = 0.466^{+0.011}_{-0.01}\mathrm{\pi}$, which corresponds to Milgrom's constant $a_0 = 2\gamma_0^{-1}c/t = 1.01_{-0.32}^{+0.33} \times 10^{-10} \mathrm{m\,s}^{-2}$ \citep{Monjo2023}. In the right panel, both the general model and the specific approach for clusters use only one free parameter ($\varepsilon_H$), done by setting the projective angle of galaxies to $\gamma_{cen} = \mathrm{\pi}/2$ and $\gamma_U = \mathrm{\pi}/3$. The general model (grey band) is represented by $\varepsilon_H = 56_{-12}^{+22}$, while the cluster approach (red band) considers an average of $\varepsilon_H = 40_{-6}^{+8}$. The shaded areas represent the 90\% confidence intervals.}
    \label{fig:Fig01}
\end{figure*}

Larger anomalies in acceleration are found for the higher orbital speeds ($v_N/v_H \sim \varepsilon_H/\sqrt{2}$) in clusters. However, this is the opposite for galaxies, which experience the maximum anomaly for low orbital velocities ($v_N/v_H < \varepsilon_H/\sqrt{2}$), as shown in Figure~\ref{fig:Fig01}. In clusters, the relative density between the dominant galaxy (BCG) and the neighborhood determines this opposite behavior. The equilibrium value of $v_N/v_H \approx \varepsilon_H/\sqrt{2}$ points to the transition regime between small and large anomalies, i.e. as small and large proportions of missing gravity using standard physics.

\subsection{Predictions for galaxy and cluster dynamics}

As discussed in Section~\ref{sec:intro}, HMG derives a relationship between the Milgromian acceleration parameter $a_0 \approx 1.2 \times 10^{-10}\text{m/s}^{2}$ and the cosmic parameter $c/t \approx 6.9 \times 10^{-10}\text{m/s}^{2}$, since $a_0 \approx a_{\gamma 0} \equiv 2\gamma_0^{-1}c/t$ for galaxy rotation curves assuming an approximately constant $\gamma_0^{-1} \approx 0.08$ \citep{Monjo2023}. However, the geometry of gravitational systems leads to a variable projection factor $\gamma_0^{-1} \in (0, 1)$, depending on the ratio between Newtonian orbital speed and Hubble flux.

According to the general model of projective angles (Equation~\ref{eq:model_general2}), it is expected that galaxies and galaxy clusters exhibit opposite behaviors in their dependence on the speed, but both follow the same theoretical curve. Using $\gamma_{cen} = \mathrm{\pi}/2$, $\gamma_U=\mathrm{\pi}/3$, and $\varepsilon_H = 56_{-12}^{+22}$ as obtained from the cluster data, we apply Equations~\ref{eq:RAR} and \ref{eq:model_general2} to predict the behavior of 60 galaxies whose data were collected by \citet{McGaugh2007}. By directly applying Equation~\ref{eq:model_general2} to the orbital speed of galaxies, the relative anomaly $\left( a_{tot} - a_N \right)/\left( c/t \right) = \gamma_0^{-1}$  is predicted to lie between 0.05 and 0.40, which is close to the value of $\gamma_0^{-1} = 0.07_{-0.02}^{+0.03}$ implied by galaxy rotation curves.

\begin{figure*}
    \centering
    \includegraphics[width=0.88\textwidth]{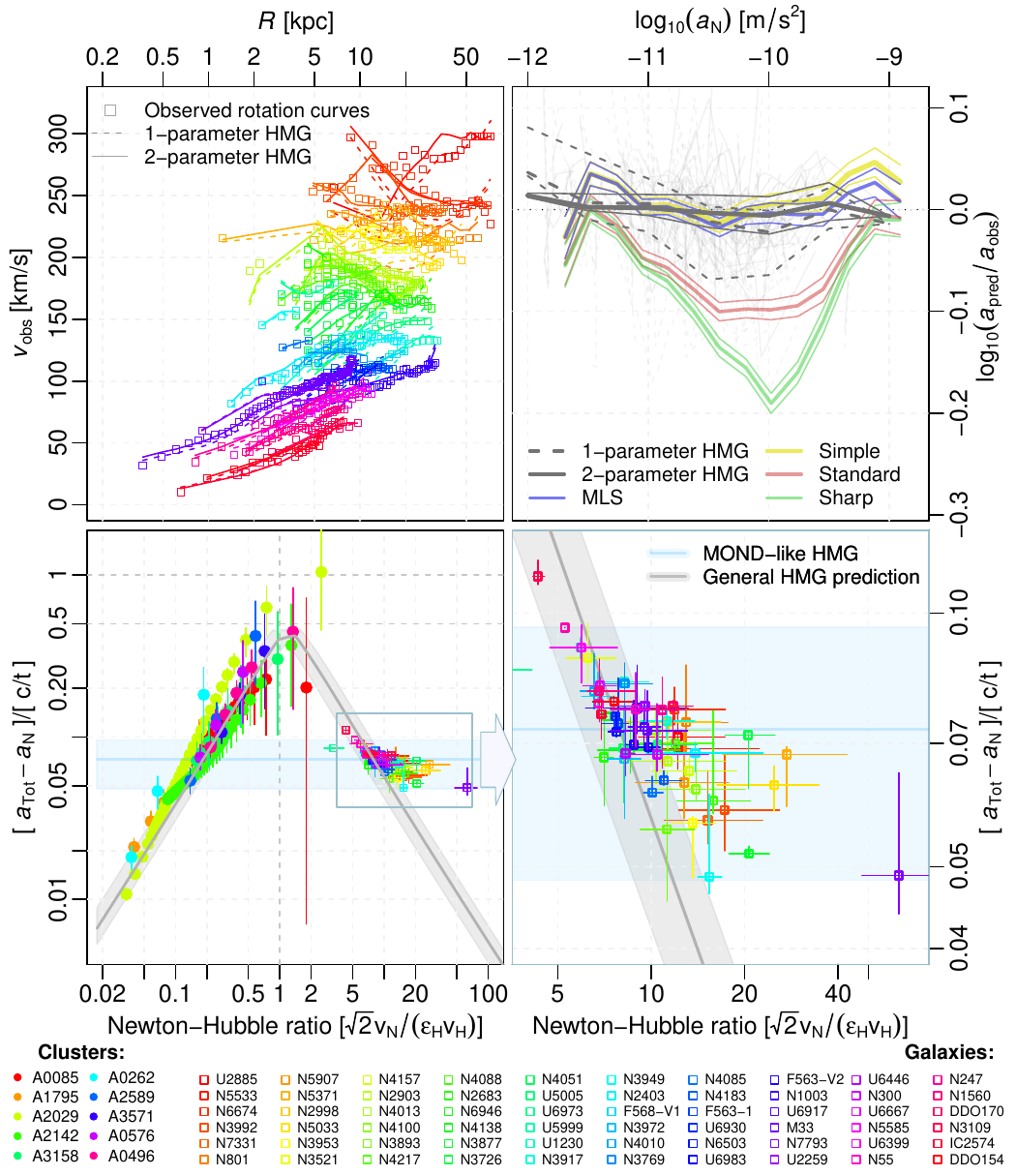}
    \caption{Galaxy rotation curves \citep[squares;][]{McGaugh2007} compared with the cluster RAR \citep[circles;][]{Li2023}. \emph{Top left}: Fitting of galaxy rotation curves according to the HMG model \citep[Equation~\ref{eq:RAR_aNc};][]{Monjo2023}, where $\gamma_0$ is modeled by Equation~\ref{eq:model_general2} with one free parameter ($\epsilon_H$, while $\gamma_{cen} = 0.48\mathrm{\pi}$; dashed lines) or two parameters ($\epsilon_H$ and $\gamma_{cen}$; solid lines). \emph{Top right}: Performance of the 1-parameter (dashed grey lines) and 2-parameter (solid grey lines) model for the ratio of predicted ($a_{pred}$) and observed ($a_{obs}$) centripetal acceleration for the 60 galaxies (light grey lines) and comparison with MOND interpolation functions (MLS, simple, standard, and sharp functions, fitted to 153 suitable galaxies). $1\sigma$ confidence intervals for all functions (upper and lower lines of each color) were found as in figure~23 of \citet{Banik2023}. \emph{Bottom left}: Global fitting of the dataset according to Equations~\ref{eq:RAR} and \ref{eq:model_general2}. \emph{Bottom right}: Zoom in on the theoretical prediction made for galaxies fitted using Equation~\ref{eq:model_general2} with $\varepsilon_H = 21_{-11}^{+32}$. As in Figure~\ref{fig:Fig01}, the general model prediction for galaxies (grey band) corresponds to the parameter $\varepsilon_H = 56_{-12}^{+22}$ as an average value of the cluster fitting, with fixed $\gamma_{cen} = \mathrm{\pi}/2$ and $\gamma_{U} = \mathrm{\pi}/3$. To compare, the MOND-like model and the observational constraint of the general HMG model are also shown, with the shaded area representing the 90\% confidence interval.}
    \label{fig:Fig02}
\end{figure*}

The wide range of $\gamma_0^{-1}$ in clusters depends on the ratio $v_N/v_H$ between the orbital speed $v_N$ and the Hubble flux $v_H$, the central projective angle $0.47\mathrm{\pi} \la \gamma_{cen} \la 0.50\mathrm{\pi}$, and the parameter $\varepsilon_H \ge 1$. For galaxy rotation curves, this additional dependency is not evident beyond the usual dependence on $a_N$ according to thorough reviews of MOND interpolating functions, showing that $\gamma_0$ is almost constant and the actual gravity only depends on the Newtonian acceleration \citep{Banik2022, Stiskalek2023}. This apparent weakness of the model is easily solved by the fact that an almost constant $\gamma_0^{-1}$ is obtained from the 2-parameter model (Equation ~\ref{eq:model_general2}) with fitted values of $\gamma_{cen}$ (top left panel of Figure~\ref{fig:Fig02}). Moreover, the relation of the rotation curve with $v_N/v_H$ is highly nonlinear as it depends on trigonometric functions. Finally, the correlation between the Newtonian acceleration $a_N \propto r^{-2}$ and the flux ratio $v_N/v_H \propto r^{-3/2}$ is very high for galaxies ($R \approx 0.90$, $p$-value $<0.001$), so the families of interpolating function $f\left(a_N\right)$ remove almost all this non-linear dependency. In any case, the effective interpolating function of the HMG model is compatible with the best MOND functions (top right panel of Figure~\ref{fig:Fig02}). It is important to note that HMG predicts the form of the interpolating function, which is arbitrary in MOND and must be found from observations.

After applying an observational constraint of Equation~\ref{eq:model_general2} to the galaxy rotation curves with $\gamma_{U} = \mathrm{\pi}/3$, we get a value of $\varepsilon_H = 21_{-11}^{+32}$ for $\gamma_{cen} = \mathrm{\pi}/2$, while assuming instead that $\gamma_{cen} = 0.48\mathrm{\pi}$ gives and $\varepsilon_H = 18_{-10}^{+28}$. Both are statistically compatible with the cluster-based fitting, which gives $\varepsilon_H = 56_{-12}^{+22}$ (bottom left panel of Figure~\ref{fig:Fig02}). In particular, a value of $\epsilon_H \approx 45$ is compatible with both datasets, albeit with a wide variability between the different cases. However, the parameter $\varepsilon_H$ is not free at all because there is a significant correlation ($R > 0.85$, $p$-value $<0.001$) of the form $\varepsilon_H \propto \sqrt{\rho}$ for galactic mass densities $\rho$ at distances between $50-200$~kpc (Appendix~\ref{sec:individual_fitting} and Figure~\ref{fig:FigC3}). This gives $\varepsilon_H \approx \sqrt{\rho/\rho_{vac}}$ for the vacuum density $\rho_{vac} = 3/\left( 8\mathrm{\pi}Gt^2 \right)$, justifying the name of the parameter $\varepsilon_H$ as the square root of the \textit{relative density of the neighborhood} (Equation~\ref{eq:model_gsys1}). Finally, an empirical relationship ($R > 0.80$, $p$-value $< 0.001$) is also found between $\cos \gamma_{cen}$ and $\log \varepsilon_H$ for galaxies, which suggests that the tight range of values of $\gamma_{cen}$ strongly depends on the geometrical features of the gravitational system.

The range of $\epsilon_H$ (usually between 10 and 100) ensures that galaxies and clusters show a flat rotation curve that extends up to at least 1~Mpc, which is quite consistent with the extended flat rotation curves around isolated galaxies as revealed by weak lensing of background galaxies collected from the KiDS survey \citep{Brouwer2021, Mistele2024}. The clusters N5044, N533, A2717, and A2029, with a baryonic mass between $10^{12} M_\odot$ and $10^{13} M_\odot$ enclosed within a radius of 1~Mpc \citep{Angus2008}, produce the following values: (1) an empirical density of $\epsilon_H \sim 10$; (2) an acceleration between $10^{-13}$ and $10^{-11}$ m/s\textsuperscript{2}; (3) a Hubble flux $v_H$ of about 70 km/s; and (4) a circular speed $v_N$ between 70 and 600 km/s. These circular speeds at 1~Mpc imply that the projection factor $\gamma_{0}^{-1}$ is between $0.07$ and $0.35$, producing anomalous accelerations equal to or greater than the Milgromian scale $a_0$ as $\gamma_0^{-1} = 0.08\pm 0.01$. The anomaly is noticeably reduced to $0.003 \le \gamma_{0}^{-1} \le 0.03$ at 10~Mpc, which is less than half of the Milgromian acceleration.

\subsection{Prediction for small systems}

For small gravitational systems, the orbital velocity $v_N$ is much higher than the Hubble flux $v_H$, so it is expected that cosmic effects are negligible (with $v_N/v_H \gg \varepsilon_H/\sqrt{2}$). This is because the ratio between the Kepler-Newton speed and Hubble flux is independent of the size of a spherical system with constant density, but smaller systems are usually much denser than larger systems. For example, according to Equation~\ref{eq:model_general2}, an anomaly of only $6.4_{-0.4}^{+1.0}\times 10^{-17}\;\text{m/s}^2$ (90\% confidence level) is predicted for the Solar System at a distance of Pluto's orbit. The predicted anomaly is even smaller for Saturn at 10~AU, which is well consistent with the null detection of anomalous effects there from Cassini radio tracking data \citep{Hees2014, Desmond2024}. For the Oort cloud, which hypothetically extends between 2 and 200~kAU, the predicted anomaly $\Delta a \equiv \gamma_0^{-1}c/t$ increases from $2.2_{-0.1}^{+0.4} \times 10^{-14}\;\text{m/s}^2$ to $2.3_{-0.1}^{+0.4} \times 10^{-11}\;\text{m/s}^2$, respectively. The latter value is about 20\% of the Milgrom acceleration $a_0 \approx 1.2 \times 10^{-10}\;\text{m/s}^2$ and could therefore be detected in the future. The most aligned finding is that shown by the work of \citet{Migaszewski2023}, who suggest that Milgromian gravity could explain the observed anomalies of extreme trans-Neptunian objects and the Oort cloud ($2-200$~kAU, up to 20\% of $a_N$). \citet{Brown_2023} claimed that the farthest Kuiper Belt objects ($\sim 250$ AU) also present a MOND signal, but the very detailed orbit integrations performed by \citet{Vokrouhlicky_2024} suggest that this interpretation neglects the crucial role of the external field direction rotating as the Sun orbits the Galaxy. Once this is included, it is clear that MOND cannot explain the clustering of orbital elements presented by observations. In fact, \citet{Vokrouhlicky_2024} exclude the possible effects of MOND on scales up to about $5-10$~kAU, which is more consistent with the findings of \citet{Migaszewski2023}. Therefore, these findings require further analysis to compare them with the hypothesis of a ninth planet in the trans-Neptunian region \citep{Batygin2024}.

In the case of wide binaries, the typical orbital speed at $r \sim 0.1$~pc is about $v_N = 350$ m/s while the Hubble flux is $v_H = r/t \sim 7\times 10^{-3}\;\text{m/s}$. Thus, the Newtonian acceleration is $a_N = v_N^2/r \sim 4.1 \times 10^{-11}\;\text{m/s}^2 < a_{0} \approx 1.2 \times 10^{-10}\;\text{m/s}^2$. This is theoretically within the classical MOND regime as $a_N < a_0$. However, the speed flux is $v_N/v_H  = 5 \times 10^4$, so we expect only a very small anomaly (i.e., a large projective angle $\gamma_{sys}$). Assuming that $\gamma_{cen}=\mathrm{\pi}/2$, $\gamma_U=\mathrm{\pi}/3$, and a global value of $\varepsilon_H = 40_{-20}^{+30}$ in Equation~\ref{eq:model_general2}, the projective angle $\gamma_{sys} = \mathrm{\pi}/2 - 2.20_{-0.35}^{+0.14} \times 10^{-4} \mathrm{\pi}$, which corresponds to a projection parameter of $\gamma_0^{-1} = \gamma_{sys}^{-1} \cos \gamma_{sys} = 4.4^{+0.7}_{-0.3} \times 10^{-4}$. The acceleration anomaly would then be $\Delta a \equiv a_{tot} - a_N = \gamma_0^{-1}c/t = 3.1^{+0.4}_{-0.2} \times 10^{-13}\;\text{m/s}^2$ (90\% interval), that is, a prediction of $2.9 \times 10^{-13}\;\text{m/s}^2 < \Delta a < 3.5 \times 10^{-13}\;\text{m/s}^2$ at the 90\% confidence level. 

Therefore, the acceleration anomalies expected for small systems are less than 1\% of the original Milgrom constant $a_0 \approx 1.2\times 10^{-10}\;\text{m/s}^2$. This result is consistent with recent comparisons between standard gravity and MOND with data from Gaia wide binary systems \citep{Pittordis2019, Pittordis2023, Banik2023}. However, some authors dispute these results by using different data selection criteria and analysis techniques \citep[e.g.][]{Hernandez_2023, Hernandez_2023b, Chae_2023, Chae_2023b, Chae_2024}).

\subsection{Strengths and limitations of the model}

The HMG model is not a complete cosmology. Our focus here is on virialised structures and dynamics in the late ($z < 2$) Universe. With a careful comparison to the observed RAR across the full range of probed accelerations in SPARC galaxies and HIFLUGCS galaxy clusters, we show that our 2-parameter model is compatible with the latest observations. Moreover, our two parameters are not completely free $-$ they naturally relate to the density of observed matter (Figure~\ref{fig:FigC3}). This is an advantage because the usual modified gravity theories have difficulties in explaining why the RAR differs between galaxies and clusters \citep[e.g., see figure~5 of][]{Li2023}.

Nevertheless, HMG faces some limitations in the early epochs of the Universe because the model distinguishes between extrinsic (linear expansion) and intrinsic (fictitious acceleration) perspectives. In particular, linear expansion cannot explain the observed cosmic microwave background (CMB) power spectrum and Big Bang nucleosynthesis (BBN) due to its very non-standard timing of key phase transitions, which occur at particular temperatures and thus values of the scale factor \citep{Faisal2019, Skordis2021, grohs2023}. Altering the timeline by even a single minute could have serious implications for BBN given that free neutrons have a decay time of 15~minutes \citep{Cyburt2016, Haslbauer2020, Banik2022}. Therefore, the intrinsic viewpoint of the HMG model needs to address earlier periods in which the expansion history is closer to the standard cosmological model.

According to \cite{MCS2023}, the apparent cosmic timeline of the intrinsic hyperconical universe is the same as the standard model, producing a fictitious acceleration compatible with the observed dark energy phenomenology \citep{Monjo2024b, Monjo2024}. In particular, the hyperconical model shows that there exists a unique local solution that produces a fictitious dark energy of $\Omega_{\Lambda} = 2/3$, while global solutions produce $\Omega_{\Lambda} \approx 0.7$. This suggests the HMG model is promising at late times. In future work, we will deeply analyze the ability of the model to reproduce both CMB and BBN observations. The key idea is the \textit{dynamical embedding technique} in which linear expansion is actually only in the ambient spacetime as a global time (geometrically, it is the transverse flux of the slicing/foliation in Cauchy surfaces). On the other hand, the redshift describes the geodesics of light in the intrinsic metric that presents fictitious acceleration, so its apparent timeline should be the standard one (as described by observers).

\section{Concluding remarks}
\label{sec:conclusions}

Acceleration is not a geometrical invariant, but rather depends on the reference system or framework considered. The hyperconical model (HMG) showed that it is possible to derive local-scale General Relativity to model gravitational systems with anomalous acceleration similar to that attributed to dark matter or dark energy \citep{MCS2023, Monjo2023}. Other MOND-based relativistic theories also obtained good performance when modeling galaxy rotation curves with a single global parameter based on acceleration. However, parameters other than acceleration are required because the classical MOND-based RAR does not extend to clusters, and also because recent observations show that gravity is mostly Newtonian on scales smaller than about 10~kAU with high precision, even at low acceleration.

This study presented a generalized applicability of the HMG model for a wide range of acceleration anomalies in gravitational systems. Good agreement was obtained with the data collected from 10 galaxy clusters and 60 high-quality galaxy rotation curves. The technique developed for the perturbed metric follows the geometric definition of the sinus of a characteristic angle $\gamma_{sys}$ as a function of the Newtonian orbital speed ($v_N$) and the Hubble flux ($\epsilon_H v_H$), i.e., $\sin \gamma_{sys}^2 - \sin^2 \gamma_U \approx \beta^2(r)\left|2v_N^2 - \epsilon^2_H v_H^2(r)\right|$ for $\gamma_U = \mathrm{\pi}/3$. The function $\beta\left( r \right)$ does not depend on the speeds. It can be fixed by setting two parameters as $\gamma_{sys} = \gamma_{sys}(\gamma_{cen}, \varepsilon_H)$. These parameters are a central projective angle $0.47\mathrm{\pi} \la \gamma_{cen} \la 0.50\mathrm{\pi}$ and a relative density $\varepsilon_H \ge 1$.

From the fitting of the general model of $\gamma_{sys}$ (Equation~\ref{eq:model_general2}) to the cluster RAR data, an anomaly between $0.05\,c/t$ and $0.40\,c/t$ is predicted for the galaxy rotation dynamics. This is statistically compatible with the observational estimate of $0.07_{-0.02}^{+0.03}c/t$. As for any modified gravity theory, the challenge was to derive a tight RAR compatible with observations with few free parameters. Classical MOND only has a global free parameter $a_0$, but it does not specify the interpolating function, so MOND actually has a lot of freedom to fit observations of galaxy dynamics. In contrast, the HMG model derives a unique interpolation function for rotation curves using only two parameters $\left( \varepsilon_H, \gamma_{cen} \right)$ that are not totally free because they are related to the density of matter.

For objects in the outer Solar System such as the farthest Kuiper Belt objects or bodies in the Oort cloud, anomalies between $10^{-14}\;\text{m/s}^2$ and $10^{-11}\;\text{m/s}^2$ are predicted at 1~kAU and at 100~kAU, respectively. Similarly, for wide binary systems, anomalies are expected within the range of $2.9 \times 10^{-13} < \Delta a < 3.5 \times 10^{-13}\;\text{m/s}^2$ (90\% confidence level). Such small predicted anomalies imply that local wide binaries should be Newtonian to high precision, as is within the observational limits.

This work provides a chance to falsify a wide range of predictions of a relativistic MOND-like theory that has previously collected successful results in cosmology \citep{Monjo2024}. In future work, we will address other open challenges, especially the modelling of cosmic structure growth and the evolution of early stages of the Universe related to the CMB angular power spectrum and the BBN observations.

\section*{Acknowledgements}

IB is supported by Science and Technology Facilities Council grant ST/V000861/1. The authors thank Prof. Stacy McGaugh for providing the data for 60 high-quality galaxy rotation curves. The data set corresponding to the 10 galaxy clusters was provided by Prof. Pengfei Li, a kind gesture which we greatly appreciate. 

\section*{Data Availability}

In this study, no new data was created or measured.

\bibliographystyle{mnras}
\bibliography{00_references}

\begin{appendix}

\numberwithin{figure}{section}

\section{Perturbed vacuum Lagrangian density}
\label{annex:A} 

\noindent 
This appendix summarizes the definition of the local Einstein field equations according to the hyperconical model, that is, by assuming that GR is only valid at local scales \citep{MCS2020,Monjo2024}. In particular, the new Lagrangian density of the Einstein-Hilbert action is obtained by extracting the background scalar curvature $R_{hyp}$ from the total curvature scalar $R \to \Delta R \equiv R - R_{hyp}$ as follows:
\begin{eqnarray} \nonumber
\mathcal{L} &=& \frac{1}{16\mathrm{\pi}G} \Delta R +
\mathcal{L}_M  = \\
&=& \frac{1}{16\mathrm{\pi}G}\left(R + \frac{6}{t^2} \right) -
\rho_M  =  \frac{c^2}{16\mathrm{\pi}G}R - \Delta \rho\,,  \label{eq:space.lagrangian3}
\end{eqnarray} 
where $G$ is the Newtonian constant of gravitation, $R_{hyp} = -{6}/{t^2}$ is the curvature scalar of the (empty) hyperconical universe, $\mathcal{L}_M = -\rho_M$ is the Lagrangian density of classical matter, and $\Delta \rho \equiv \rho_M - \rho_{vac}$ is the density perturbation compared to the `vacuum energy' $\rho_{vac} = 3/(8\mathrm{\pi}Gt^2)$ with mass-related event radius $r_M \equiv 2G\widetilde{M} \equiv 2G\rho_{vac} \, \frac{4}{3} \mathrm{\pi} t^3 = t$, where $\widetilde{M}$ is a `total mass' linked to $\rho_{vac}$. Moreover, the orbital velocity $v_N(\rho_{vac})$ associated with $\rho_{vac}$ at $r$ is given by $2v_N^2(\rho_{vac}) =  2G \rho_{vac} \, \frac{4}{3} \mathrm{\pi} r^3 = r^2/t^2 = v_H^2$. Therefore, a total density $\rho_M$ leads to a total squared orbital velocity $v_N^2(\rho_M)$ as follows:
\begin{eqnarray} \nonumber
2v_N^2(\rho_M) &=& 2G \rho_M \, \frac{4}{3} \mathrm{\pi} r^3 = 2G \left(\rho_{vac} + \Delta\rho  \right)\, \frac{4}{3} \mathrm{\pi} r^3 = 
\\
&=& \frac{r^2}{t^2} + \frac{2GM}{r} = 2v_N^2(\rho_{vac}) + 2v_N^2(\Delta\rho)\,, \label{eq:vacuum}
\end{eqnarray}
where we use the definition of $M \equiv \Delta\rho \, \frac{4}{3} \mathrm{\pi} r^3$. Now, let $\theta_M \equiv M/\widetilde{M} \ll 1$ be a (small) constant fraction of energy corresponding to the perturbation $\Delta \rho$, and $r_M \equiv 2GM = \theta_M t$ be the radius of the mass-related event horizon. Thus,
\begin{eqnarray}
\frac{2GM}{r} = \frac{\theta_M t}{r'\frac{t}{t_0}} = \frac{\theta_M t_0}{r'} =: \frac{2 GM_0}{r'}\,.
\end{eqnarray}
Therefore, the quotient $M/r = M_0/r'$ is as comoving as $r/t = r'/t_0$.

Moreover, the background metric of the universe has a Ricci tensor with components $R_{00}^u = 0$ and $R_{ij}^u = \frac{1}{3} {R}_u \, g_{ij}$ \citep{Monjo2017,MCS2020}. Since $R_{hyp} = -6/t^2$, the Einstein field equations become locally converted to \citep{Monjo2024}: \begin{eqnarray}\begin{cases} \label{eq:modified_EFE1}
   \kappa P_{00} & =   \Delta R_{00} - \frac{1}{2}\Delta R g_{00} \;\; =  \;\;  R_{00} - \frac{1}{2} R g_{00} - \frac{3}{t^2}g_{00} \\ 
\kappa P_{ij} & =   \Delta R_{ij} - \frac{1}{2}\Delta R g_{ij} \;\; =  \;\; R_{ij} - \frac{1}{2} R g_{ij} - \frac{1}{t^2} g_{ij}
\end{cases}\,,
\end{eqnarray} where $\kappa = 8\mathrm{\pi}G$ and $P_{\mu\nu}$ are the stress-energy tensor components. Notice that, for small variations in time $\Delta t = t - t_0 \ll t_0 \equiv 1$, the last terms ($3/t^2$ and $1/t^2$) are equivalent to consider a `cosmological (almost) constant' or dark energy with equation of state $w = -1/3$ (varying as $a^{-2}$).

\section{Hyperconical modified gravity (HMG)}
\label{sec:hyperconical}

\subsection{Hyperconical universe and its projection}
\label{sec:projection}

This appendix reviews the main features of relativistic MOND-like modified gravity derived from the hyperconical model and referred to here as HMG \citep{Monjo2017, Monjo2018, MCS2020, MCS2023, Monjo2023}. Let ${\pazocal{H}^4}$ be a (hyperconical) manifold with the following metric: \begin{equation} \label{eq:hyp_metric}
g_{hyp} \approx dt^2 (1- kr'^2 ) 
-  \frac{t^2}{t_{0}^2} \left( \frac{dr'^2}{1-kr'^2} + {r'}^2d{\Sigma}^2 \right)
-  \frac{2r'tdr'dt}{t_{0}^2\sqrt{1-kr'^2}}\,, 
\end{equation}where $k = 1/t_0^2$ is the spatial curvature for the current value $t_0 \equiv 1$ of the age $t$ of the universe, while ${a}(t) \equiv t/{t_0}$ is a linear scale factor, $r' \ll t_0$ is the comoving , and $\Sigma$ represents the angular coordinates. Both the (Ricci) curvature scalar and the Friedmann equations derived for $k=1$ are locally equivalent to those obtained for a spatially flat ($K_{FLRW} = 0$) $\Lambda$CDM model with linear expansion \citep{MCS2020}. In particular, the local curvature scalar at every point ($r'\equiv 0$) is equal to \citep{Monjo2017}:\begin{eqnarray}
    R_{hyp} = -\frac{6}{t^2} =  R_{FLRW}\bigg|_{K\,=\, 0,\; a\,=\,t/t_0}\,,
\end{eqnarray}as for a three-sphere (of radius $t$). This is not accidental because, according to \citet{MCS2020}, the local conservative condition in dynamical systems only ensures internal consistency for $k=1$.


The hyperconical metric (Equation~\ref{eq:hyp_metric}) has shift and lapse terms that produce an apparent radial inhomogeneity, which is equivalent to an acceleration. This inhomogeneity can be assimilated as an apparent acceleration by applying some `flattening' or spatial projection. In particular, for small regions, a final intrinsic comoving distance $\hat{r}'$ can be defined by an $\alpha$-distorting stereographic projection \citep{Monjo2018, MCS2023},	
\begin{eqnarray}
     \label{eq:projectionmap} 
     r' & \mapsto & \hat{r}' = \frac{r'}{\left(1-\frac{\gamma(r')}{\gamma_0}\right)^\alpha}  \,,\\
         t & \mapsto & \hat{t} =  \frac{t}{1-\frac{\gamma(r')}{\gamma_0}}  \,,
\end{eqnarray}
where $\gamma = \gamma(r') \equiv \sin^{-1}(r'/t_0)$ is the angular comoving coordinate, $\gamma_0^{-1} \in (0, 1)$ is a projection factor, and $\alpha = 1/2$ is a distortion parameter, which is fixed according to symplectic symmetries \citep{MCS2023}. Locally, for empty spacetimes, it is expected that $\gamma_0 \approx 2$; which is compatible with the fitted value of $\gamma_0 = 1.6^{+0.4}_{-0.2}$ when Type Ia SNe observations are used \citep{MCS2023}. In summary, the projection factor $\gamma_0$ depends on a projective angle $\gamma_{sys}$ such that $\gamma_0 = \gamma_{sys}/\cos(\gamma_{sys}) \ge 1$, where $\gamma_0 = 2$ corresponds to a total empty projective angle of $\gamma_{sys} \approx \mathrm{\pi}/3$, and $\gamma_0 = 1$ is the minimum projection angle allowed by the causality relationship of the arc length $\gamma_0 t_0$. Therefore, the projective angle for an empty or almost empty neighborhood is approximately $\gamma_{neigh} = (0.284 \pm 0.049)\mathrm{\pi} \la \mathrm{\pi}/3 =: \gamma_U$. 

\subsection{Perturbation by gravitationally bound systems}

In the case of an (unperturbed) homogeneous universe, the linear expansion of ${\pazocal{H}^4}$ can be expressed in terms of the vacuum energy density $\rho_{vac}(t) = 
3/(8\mathrm{\pi}Gt^2)$, where $G$ is the Newtonian gravitational constant, and thus $\rho_{vac}(t_0) = \rho_{crit}$. That is, one can define an inactive (vacuum) mass or energy $\mathcal{M}(r) = \rho_{vac} \frac{4}{3}\mathrm{\pi} r^3$ for a distance equal to $r$ with respect to the \textit{reference frame origin}. Using the relationship between the coordinate $r$ and the comoving $r'$, the spatial dependence of the metric is now \begin{eqnarray}\label{eq:metric_GM}
\frac{{r'}^2}{t_{0}^2} = \frac{{r}^2}{t^2} = \frac{2G\rho_{vac}\frac{4}{3}\mathrm{\pi} r^3}{r}  = \frac{2G\mathcal{M}(r)}{r} = v_H^2(r) \,,
\end{eqnarray} where $v_H(r) \equiv r/t$ is the Hubble speed, which coincides with the escape speed of the empty spacetime with vacuum density $\rho_{vac}$. 

\begin{definition}[\textbf{Mass of perturbation}] A perturbation of the vacuum density $\rho_{vac} \to \rho_M(r) \equiv \rho_{vac} + \Delta \rho$, with an effective density $\Delta \rho$ at $r > 0$, leads to a system mass $M_{sys} \equiv \frac{4}{3}\mathrm{\pi} r^3 \Delta \rho$ enclosed at a radius $r$, which is likewise obtained by perturbing the curvature term,\begin{eqnarray}\label{eq:t_sys}
    \frac{r^2}{t^2} \to \frac{r^2}{t_{sys}(r)^2} \equiv \frac{r^2}{t^2} + \frac{2GM_{sys}}{r} = v_H^2(r) + 2v_N^2(r)\,,
\end{eqnarray} with a radius of curvature $t_{sys}(r) \in (2GM_{sys}, t]$, where $v_N(r) \equiv \sqrt{GM_{sys}/r}$ is the classical Kepler-Newton orbital speed (Equation~\ref{eq:vacuum}).\end{definition}

An approximation to the Schwarzschild solution can be obtained in a flat five-dimensional ambient space from the hyperconical metric. For example, let $(t, \vec{r}, u) \equiv (t, x, y, z, u) \in \mathbb{R}_{\eta}^{1,4}$ be Cartesian coordinates, including an extra spatial dimension $u$ in the five-dimensional Minkowski plane. As used in hyperconical embedding, $u \equiv  t \cos\gamma  - t$ is chosen to mix space and time. Now, it includes a gravity field with system mass $M_{sys}$ integrated over a distance $\hat r$ such that $\sin^2 \gamma \equiv \frac{r^2}{t^2} \mapsto \frac{\hat r^2}{t^2} + \frac{2G{M_{sys}}}{\hat r}$. Notice that $r$ is a coordinate related to the position considered, in contrast to the observed radial distance $\hat r$ or its comoving distance $\hat r' \equiv (t_0/t) \hat r$. With this, first-order components $\hat g_{\mu\nu}$ of the metric perturbed by the mass are:\begin{eqnarray} \nonumber
   \hat  g_{tt} &=&  2\cos\gamma - 1 \;\; \approx   \;\; 1 - \frac{{\hat r}^2}{t^2} - \frac{2GM_{sys}}{{\hat r}}\,,
     \\ \nonumber
\hat g_{r'r'}  &=&  - \frac{t^2}{t_0^2} \frac{1}{\cos^2\gamma} 
\; = \;
- \frac{t^2}{t_0^2} \left(1 - \frac{{\hat r}^2}{t^2} - \frac{\mathrm{2G}{M_{sys}}}{{\hat r}} \right)^{-1} 
\approx  
\\ \nonumber
&&\; \;  \approx
- \frac{t^2}{t_0^2} \left(1 + \frac{{\hat r}^2}{t^2} + \frac{\mathrm{2G}{M_{sys}}}{{\hat r}} \right)\,,
\\ \nonumber
\hat g_{r't}  &=& \frac{t}{t_0}\tan\gamma \;\;  =\;\;  \frac{t}{t_0}\frac{\hat r}{t} \left(1 - \frac{{\hat r}^2}{t^2} - \frac{2G{M_{sys}}}{{\hat r}} \right)^{-1/2} \;  \approx  
\\ \nonumber
&&\; \;  \approx \frac{t}{t_0}\frac{\hat r}{t} + O\left(\frac{{\hat r}^3}{2t^3}\right)\,,
\\ \nonumber
\hat g_{\theta\theta} &=& - \frac{t^2}{t_0^2} {\hat r}'^2\,,
\\ \nonumber
\hat g_{\varphi\varphi} &=& - \frac{t^2}{t_0^2} {\hat r}'^2 \sin^2\theta\,,
\end{eqnarray}where the hyperconical model is recovered taking $M_{sys} = 0$. Therefore, assuming linearized perturbations of the metric $\hat g_{\mu\nu} =  \hat g_{\mu\nu}^{back} + \hat h_{\mu\nu}$ with $\hat g_{\mu\nu}^{back} \equiv \hat g_{\mu\nu}|_{M_{sys}=\,0}$, we can find a local approach to the Schwarzschild metric perturbation $h|_{Schw}$ as follows \citep{Monjo2023}:\begin{eqnarray} \nonumber 
     &  {\hat g}_{Schw}  :\approx   \left[\eta_{\mu\nu} + \left(\hat g_{\mu\nu} - \hat g_{\mu\nu}^{back} \right)\right] dx^\mu dx^\nu   \approx 
   \\ \nonumber
  {}&  \;\; \approx   \left(1- \frac{2G{M_{sys}}}{\hat r} \right) d{\hat t}^2
- \frac{t^2}{t_{0}^2} \left[\left(1 + \frac{2G{M_{sys}}}{\hat r}\right) d{{\hat r}'^2} + {{\hat r}'^2}d{\Sigma}^2 \right] \,+ 
\\ 
 &  \;\; \;\;  + \;\;   \text{neglected shift}\,,
  \label{eq:Schwarzschild_g}
\end{eqnarray}  which is also obtained for $\hat g_{\mu\nu}$ when $\hat r/t \ll 1$, that is $\lim_{(\hat r/t_0) \to 0}  [\hat g_{\mu\nu}] \approx {\hat g}_{Schw}$. The shift term is neglected in comparison to the other terms, especially for geodesics. Our result is aligned to the Schwarzschild-like metric obtained by \citet{Mitra2014} for FLRW metrics, specifically for the case of $K=0$.

In summary, the first-order approach of the 5-dimensionally embedded (4-dimensional) hyperconical metric (Equation~\ref{eq:Schwarzschild_g}) differs from the Schwarzschild vacuum solution by the scale factor $t^2/t_0^2$ and by a negligible shift term. Therefore, the classical Newtonian limit of GR is also recovered in the hyperconical model, because the largest contribution to gravitational dynamics is given by the temporal component of the metric perturbation $h_{tt}$. That is, the Schwarzschild geodesics are linearized by
\begin{eqnarray}
\label{eq:geodesic0}
\frac{d^2x^\mu}{d\tau^2} \approx \frac{1}{2}\eta^{\mu\nu} \frac{\partial}{\partial x^\nu}  h_{tt} \left(\frac{dt}{d\tau}\right)^2 \;,
\end{eqnarray} where $\hat h_{tt} \approx - 2GM_{sys}/{\hat r}$.

\section{Modeling radial acceleration}

\label{sec:gravitational_angle}

\subsection{Projective angles of the gravitational system}
\label{sec:projective_angle}

The last appendix derived a general expression for the anomalous RAR expected for any gravitational system according to the projective angles (which depend on the quotient between orbital speed and Hubble flux) under the hyperconical universe framework. 

\begin{figure*}
\centering 
	\includegraphics[scale=0.74]{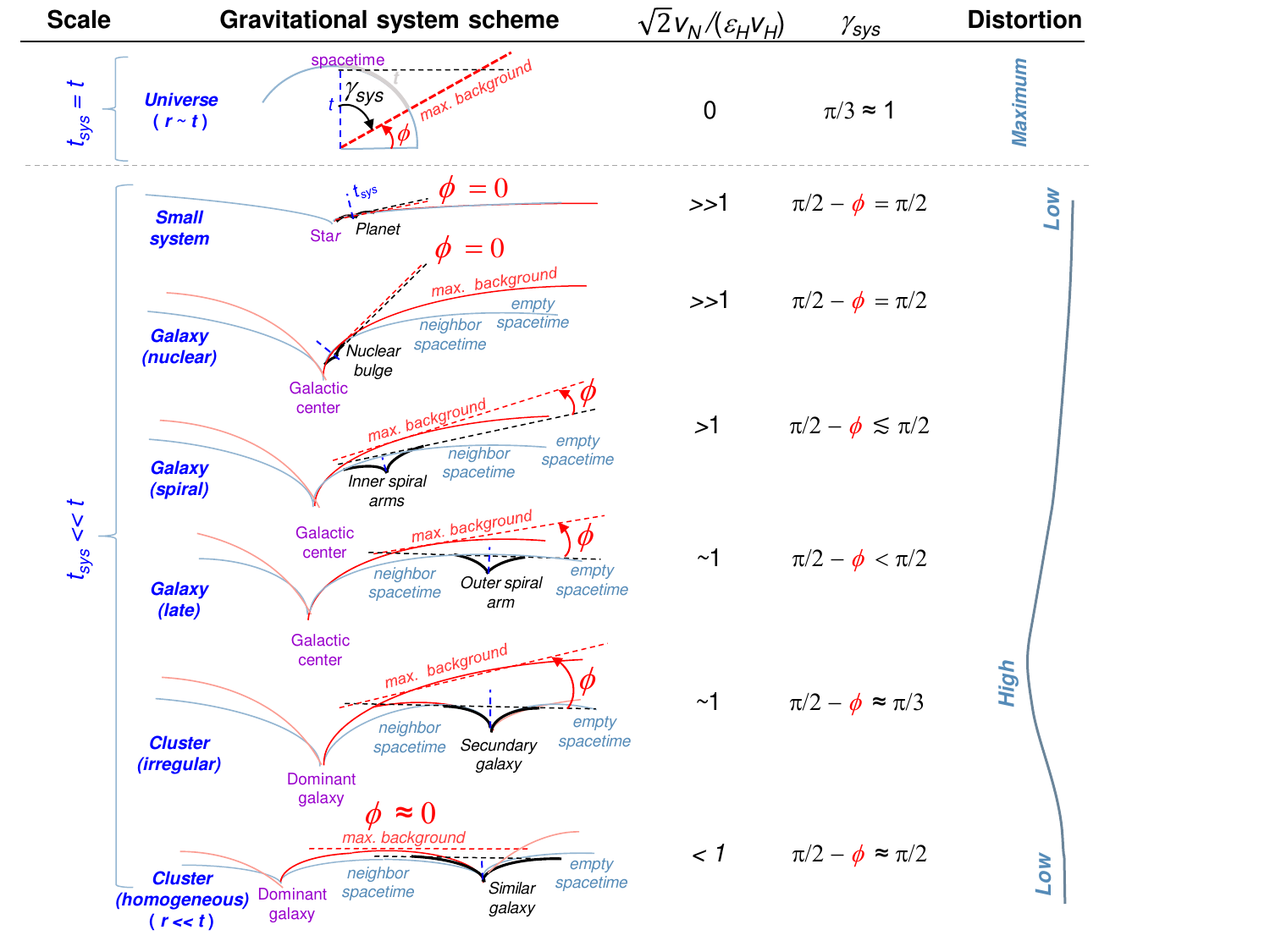}
    \caption{Conceptual model of the projective angle $\gamma_{sys}$ as a function of the relative geometry (orbital speed $v_N$ over the Hubble flux $\varepsilon_Hv_H$), for each gravitational system (black curves), with respect to the maximum background spacetime (red curves). To simplify the scheme, the projection factor $\gamma_0^{-1} \equiv \gamma_{sys}^{-1}\cos \gamma_{sys}$ is graphically represented by an auxiliary angle $\phi$, defined as the arc between the hyperplane of the gravitational system (dashed black line) and the maximum background hyperplane (dashed red line) curved by the dominant system (purple), such that $\gamma_{sys} = \mathrm{\pi}/2 - \phi$. The distance scale $r$ is represented by $t_{sys}$ (Equation~\ref{eq:t_sys}). In addition to the cosmic scale, six gravitational systems are considered: i) a small system (e.g. Solar System); ii) an early-type galaxy with a dominant nuclear bulge; iii) a typical spiral galaxy; iv) a late-type spiral galaxy; v) an irregular cluster; and vi) a homogeneous cluster. The dominant objects (purple text) are the star for a small system, the galactic center for galaxies, and the brightest galaxy for a given cluster.}  \label{fig:FigC1}
\end{figure*}

From the analysis of perturbations (Equation~\ref{eq:t_sys}), it is expected that any gravitational system (Equation~\ref{eq:Schwarzschild_g}) results in a characteristic scale $r_{cs}(M_{sys}(r)) \equiv t_{sys}(r) \sin \gamma_{sys}(r)$ given by a projective angle $\gamma_{sys} \in [\mathrm{\pi}/3, \mathrm{\pi}/2)$ that slightly depends on the radial distance $r$ and on the mass $M_{sys}$. Unlike gravitational lensing, a non-null cosmic projection $\gamma_0^{-1} = \gamma_{sys}^{-1} \cos \gamma_{sys} > 0$ is expected for non-concentrated gravitational systems. In particular, we assume that the maximum projective angle ($\gamma_{sys} = \gamma_{cen} \equiv \mathrm{\pi}/2$, minimum cosmic projection) is produced by small, dense, and homogeneous gravitational systems, while the minimum angle ($\gamma_{sys} = \gamma_{U} \equiv \mathrm{\pi}/3$, maximum cosmic projection) corresponds to large systems extended towards an (almost) empty universe (Figure~\ref{fig:FigC1}). Since $t_{sys}^2(r) \in (4G^2M_{sys}^2, t^2]$ and $r_{cs}^2(M) \in (4G^2M_{sys}^2, \frac{3}{4}t^2]$, the characteristic scale $\,r_{cs}^2(M_{sys})$ increases from $r_{cs}^2(M_{sys})=t_{sys}^2(r)=4G^2M_{sys}^2 \ll t^2$ up to $r_{cs}^2(M_{sys})=\frac{3}{4}t_{sys}^2(r)=\frac{3}{4}t^2 = t^2\sin^2\gamma_U$; that is, $\sin \gamma_{sys}^2(r) \in [\frac{3}{4},1]$. 

The relation of $\gamma_{sys}$ with respect to the gravitational mass and the scale of speeds can be estimated from the following properties. According to Equation~\ref{eq:t_sys}, a gravitational system perturbs the cosmological geometry with (squared) orbital speed of $v_N^2(r) \equiv \frac{GM_{sys}}{r}$ higher than the Hubble expansion speed $v_H^2(r) \equiv \frac{r^2}{t^2}$, so the projective angle $\gamma_{sys}$ is given by: \begin{eqnarray}  \nonumber
    \sin^2 \gamma_{sys}(r) & = & \frac{r_{cs}^2(M_{sys})}{t_{sys}^2(r)}  =  r_{cs}^2(M_{sys})\left(\frac{1}{t^2} + \frac{2GM_{sys}}{r^3}\right) \approx
    \\  \nonumber
    & \approx &  \;\; \sin^2 \gamma_{neigh}  + 2\beta^2(r) v_N^2    \; \sim
    \\  
    & \sim& 
    \sin^2 \gamma_U - \beta^2(r)\epsilon^2_H v_H^2(r)  + 2\beta^2(r) v_N^2\,,
    \label{eq:model_gsys1}  {}
\end{eqnarray} where $\beta^2(r) \equiv r_{cs}^2/r^2 \gg 1$ is an auxiliary function, $\gamma_{neigh} \equiv  \sin^{-1}(r_{cs}/t) \in (0, \gamma_U)$ is a characteristic \textit{neighbor angle}, and $\epsilon^2_H \equiv  \sin^2 \gamma_U /  \sin^2 \gamma_{neigh} - 5/6 \propto t^2/r_{cs}^2 \propto \rho/\rho_{vac}$ is a \textit{relative density} of the neighborhood matter ($\rho$) with respect to the vacuum density ($\rho_{vac}$; Equation~\ref{eq:vacuum}), which is interpreted as an equilibrium point between the orbital speed and the Hubble flux. So, roughly speaking, it is $\sin^2 \gamma_{neigh}(r) \sim \frac{5}{6}\sin^2 \gamma_{neigh}(r) = \sin^2 \gamma_U - \epsilon^2_H r_{cs}^2/t^2$. On the other hand, the center of the gravitational system presents a higher density, thus the cosmic projection should be minimum due to the maximum projective angle $\gamma_{cen} \approx \mathrm{\pi}/2$, that is, \begin{eqnarray} \nonumber
   1 & \approx  & \sin^2 \gamma_{cen} \approx \frac{r_{cs}^2(M_{sys})}{t_{sys}^2(r)} + 2\epsilon_H^2 \frac{r_{cs}^2(M_{sys})}{t^2} \sim \\
 & \sim  & \sin^2 \gamma_U + \beta^2(r)\epsilon^2_H v_H^2 + 2\beta^2(r)v_N^2
   \,. \label{eq:model_gsys2}
\end{eqnarray}Notice that, for the limit when  $\gamma_{neigh} \approx \gamma_U$, it is required that $\epsilon^2_H \approx  \frac{1}{6}$ and $v_N \approx 0$. The dependency of $\gamma_{sys}$ on the auxiliary function $\beta(r)$ can be removed by taking the quotient of $\sin^2 \gamma_{sys}(r) - \sin^2 \gamma_{U}$ (Equation~\ref{eq:model_gsys1}) over $\sin^2\gamma_{cen} -  \sin^2 \gamma_{U}$ (Equation~\ref{eq:model_gsys1}). Therefore, it is expected that the projective angle $\gamma_{sys}$ of every gravitational system presents a \textbf{general relation} similar to \begin{eqnarray} \label{eq:model_generalb}
 \frac{\sin^2 \gamma_{sys}(r) - \sin^2 \gamma_{U}}{\sin^2\gamma_{cen} -  \sin^2 \gamma_{U}} \sim \bigg| \frac{2v_N^2(r)-\epsilon^2_H v_H^2(r)}{2v_N^2(r) + \epsilon^2_H v_H^2(r)}\bigg|\,,
\end{eqnarray} with two free parameters, $\epsilon_H \ge \frac{1}{6}$ and $\gamma_{cen} \approx  \frac{1}{2}\mathrm{\pi}$. Notice that the maximum projective angle (i.e. $\gamma_{sys} = \gamma_U$) is found when the gravitational and Hubble fluxes are in equilibrium. On the other side, for galaxies and small gravitational systems, the Kepler-Newton orbital speed $v_N^2$ is very larger to the Hubble flux. Thus, the projective angle $\gamma_{sys}(r) \equiv \gamma_{gal}(r)$ can be estimated by the following \textbf{galactic relation} \citep{Monjo2023}: \begin{eqnarray} \label{eq:model_galb}
 \frac{\sin^2 \gamma_{sys}(r) - \sin^2 \gamma_{neigh}}{\sin^2\gamma_{cen} - \sin^2 \gamma_{U}} \sim \frac{2v_N^2(r)}{2v_N^2(r) + \epsilon^2_H v_H^2(r)}\,,
\end{eqnarray}with $\gamma_{neigh} \sim \gamma_U \approx \mathrm{\pi}/3$ and one free parameter, which is $\epsilon^2_H \ga 1$ if $\gamma_{cen} = \mathrm{\pi}/2$ is fixed, or $\gamma_{cen} \la \mathrm{\pi}/2$ if $\epsilon_H = 1$ is fixed. Thus, two limiting cases are $\sin \gamma_{sys} \approx 1 \, \Rightarrow\,\gamma_{sys} \approx \mathrm{\pi}/2$ when orbital speed is $v_N(r) \gg \varepsilon_H v_H(r)$, while $\sin \gamma_{sys} \approx \frac{\sqrt{3}}{2} \, \Rightarrow\,\gamma_{sys} \approx \mathrm{\pi}/3$ when orbital speed is $v_N(r) \ll \varepsilon_H v_H(r)$, which is the lower limit of the neighborhood projective angle ($\gamma_{neigh}$).

On the other hand, radial accelerations (without regular orbits) of large-scale objects such as galaxy clusters are expected to present opposite behavior with respect to Equation~\ref{eq:model_galb}, since the gravitational center is not a galactic black hole but is close to a dominant galaxy \citep[the brightest cluster galaxy (BCG);][]{Shi2023,DePropris2020}, and the neighborhood now corresponds to the large-scale environment of the clusters themselves. Therefore, the projective angle $\gamma_{sys}$ of the largest structures is approximated by the following \textbf{cluster relation}:\begin{eqnarray} \label{eq:model_clustb}
 \frac{\sin^2 \gamma_{cen} - \sin^2 \gamma_{sys}(r)}{\sin^2\gamma_{cen} -  \sin^2 \gamma_U} \sim \frac{2v_N^2(r)}{2v_N^2(r)+\epsilon^2_H v_H^2(r)}\,,
\end{eqnarray}where $v_N^2(r)$ is the orbital speed profile of the cluster, $\gamma_{cen} \la \mathrm{\pi}/2$ is the averaged projective angle for its central object and, now, we expect that the projective angle for clusters is a variable $\gamma_{sys}(r) \in [\mathrm{\pi}/3, \mathrm{\pi}/2)$, but close to the neighborhood value $\gamma_{neigh} \sim \gamma_U = \mathrm{\pi}/3$.

However, a perfectly homogeneous distribution of low-density galaxies in a cluster will lead to a balance between the different galaxies that form it, so the cluster radial acceleration will be approximately zero ($v_N \sim 0$) and anomalies are not expected, thus the projective angle will be $\gamma_{sys} \approx \mathrm{\pi}/2$ for both Equation~\ref{eq:model_generalb} and Equation~\ref{eq:model_clustb}; that is, no significant geometrical differences are expected between the external and internal parts of the cluster (see the last case of Figure~\ref{fig:FigC1}). Conversely, for irregular clusters ($v_N \sim \epsilon_Hv_H$ with $\epsilon_H \gg 1$ in Equation~\ref{eq:model_generalb} or $2v_N^2 \ga \epsilon^2_H v_H^2(r)$ in Equation~\ref{eq:model_clustb}), the radial acceleration will be very similar to the cosmic expansion (with angle $\gamma_{sys} = \gamma_U = \mathrm{\pi}/3$). Notice that, for very inhomogeneous systems ($v_N \gg \epsilon_H v_H$), Equation~\ref{eq:model_generalb} recovers the behavior of high-density galaxies (Equation~\ref{eq:model_galb}) with $\gamma_{sys}(r) = \gamma_{gal}(r)$. Moreover, for $2v_N \in (0,\; \epsilon_H v_H)$, Equation~\ref{eq:model_generalb} behaves in a similar way as in Equation~\ref{eq:model_clustb} as expected.


\subsection{Cosmological projection of the Schwarzschild metric}

Henceforth, the constant of light speed $c\equiv 1$ will not be omitted from the equations so we can compare with real observations later. Let $\lambda$ be the scaling factor of an $\alpha$-distorting stereographic projection (Equation~\ref{eq:projectionmap}) of the coordinates $(r', u) = (ct\sin\gamma,\; ct\cos\gamma) \in \mathbb{R}^2$, used to simplify the spatial coordinates $({\vec r}', u) \in \mathbb{R}^4$ due to angular symmetry. For nonempty matter densities, we contend that $\gamma_{sys}$ depends on the orbital speed of the gravity system considered. However, the first-order projection can be performed by assuming that the dependence on distances is weak (i.e., with $\gamma_0^{-1} = \gamma_{sys}^{-1} \cos \gamma_{sys}$ being approximately constant for each case). Thus, the stereographic projection is given by the scale factor $\lambda$ such as \citep[see for instance][]{MCS2023,Monjo2023}: \begin{eqnarray} \label{eq:lambda_globa2}
\lambda = \frac{1}{1-\frac{\gamma}{\gamma_0}}\approx  {1+\frac{r'}{\gamma_0 t_0c}} \,,
\end{eqnarray}where $\gamma = \sin^{-1} [r'/(t_0 c)] \approx r'/(t_0 c)$ is the angular position of the comoving distance $r' = (t_0/t)\,r$. Therefore, the projected coordinates are \begin{eqnarray} \begin{cases}
    \hat r'  =  \lambda^{\alpha} r' \approx \left( {1+\frac{\alpha r'}{\gamma_0 t_0c}} \right) r\,, \\
     \hat t = \lambda t \approx \left(1 +  \frac{r'}{\gamma_0t_0c}\right) t\,,
\end{cases}\label{eq:lambda_globa3}
\end{eqnarray} At a local scale, the value of $\alpha = 1/2$ is required to guarantee consistency in dynamical systems \citep{MCS2023}, but the parameter $\alpha$ is not essential in this work, since only the temporal coordinate is used in our approach below.

Applying this projection to the perturbed metric (Equation~\ref{eq:Schwarzschild_g}) and obtaining the corresponding geodesics, it is easy to find a first-order approach of the cosmic contribution to modify the Newtonian dynamics in the classical limit, as shown below (Sec.~\ref{sec:fist_pert_geod}).

 \subsection{First-order perturbed geodesics}
\label{sec:fist_pert_geod}

Assuming that the projection factor $\gamma_0^{-1} = \gamma_{sys}^{-1} \cos \gamma_{sys}$ is approximately constant, the quadratic form of the projected time coordinate (Equation~\ref{eq:lambda_globa3}) is as follows:
\begin{eqnarray} \label{eq:lambda_globa4} 
d \hat t^2
 \approx  \left(1 +  \frac{2 r'}{\gamma_0 t_0c} + \frac{2t\dot r'}{\gamma_0 t_0c}  \right) dt^2 \;\;+\;\;\text{higher-order terms}\,.
\end{eqnarray} By using these prescriptions, our Schwarzschild metric (Equation~\ref{eq:Schwarzschild_g}) is expressed in projected coordinates $(\hat{t}, \hat{r}')$ or in terms of the original ones $(t, r')$; that is, ${\hat g}_{Schw} = \hat{g}_{\mu\nu} d\hat{x}^\mu d\hat{x}^\nu = {g}_{\mu\nu} d{x}^\mu d{x}^\nu$, with \begin{eqnarray} \nonumber
 {\hat g}_{Schw} & \approx &  \left(1- \frac{2GM_{sys}}{c^2 \hat r} \right) c^2 d \hat t^2
-  \frac{ \hat t^2}{t_{0}^2} \, { \hat r'^2}d{\Sigma}^2  \approx
\\ \nonumber
  \;\;\;\;  & \approx & \;\;   g_{tt}c^2dt^2 +  g_{ii}(dx^i)^2 \,,
\end{eqnarray} and finally, it is locally expanded up to first-order perturbations in terms of $\gamma_0$. Notice that, according to Equation~\ref{eq:Schwarzschild_g}, the background terms $r'^2/t_0^2$ do not produce gravitational effects and thus they can be neglected. Here, one identifies a projected perturbation $h_{tt}$ of the temporal component of the metric, $g_{tt} = \eta_{tt} + h_{tt} = 1 + h_{tt}$, with $\eta_{\mu\nu} = \eta^{\mu\nu}= \mathrm{diag}(1,-1,-1,-1)$. Thus, if $M_{sys}$ is assumed to be mostly concentrated in the central region of the gravitional system, the first-order perturbation of the temporal component of the metric is \begin{eqnarray}\label{eq:gtt}
h_{tt} \approx  - \frac{2GM_{sys}}{rc^2}\left({1-\frac{\alpha r}{\gamma_0 t c}} \right)  +  \frac{2}{\gamma_0 c} \left(\frac{r}{t} + \frac{t}{ t_0} {\dot r'} \right)\,,
\end{eqnarray} where the spatial projection $\hat r \approx (1+\alpha r / (\gamma_0 ct))\, r$ is considered (from Equation~\ref{eq:lambda_globa3}), and the relation between comoving distance $r'$ and spatial coordinate $r$ is also used ($r'/t_0 = r/t$). 


Under the Newtonian limit of GR, the largest contribution to gravity dynamics is given by the temporal component of the metric perturbation $h_{tt}$. That is, Schwarzschild geodesics (Equation~\ref{eq:geodesic0}) produce both time-like and space-like acceleration components from the metric perturbation $h_{tt}$, \begin{eqnarray} 
\frac{d^2 \hat s}{c^2 dt^2} \approx \frac{1}{2} \frac{\partial}{\partial x^0} h_{tt} \,e_t \; - \;  \frac{1}{2} \frac{\partial}{\partial x^i} h_{tt}\,e_i  =: a^t e_t + a^ie_i\,,
\end{eqnarray} where the four-position $\hat s \equiv (c\Delta t, x^i) = c \Delta t\, e_t \,+\, x^i\, e_i =: c \Delta\mathbf{t} + \mathbf{x} \in \mathbf{R}^{1,3}$ is assumed, with canonical basis $\{e_t, e_1, e_2, e_3\}$ and dual basis $\{e^t, e^1, e^2, e^3\}$. For a freely falling particle with central-mass reference coordinates $\mathbf{x} = x^ie_i = (r, 0, 0) = \mathbf{r} \in \mathbb{R}^3$, it experiences an acceleration of about\begin{eqnarray} \nonumber
\frac{d^2 \hat s}{dt^2} &= &a^te_t + a^re_r \approx \;
\\ \nonumber
& \approx & \left(\frac{\dot r'}{t_0}- \frac{r + \frac{1}{2}\alpha r_M}{\gamma_0t^2}\right) e_t \; - \;  \left(\frac{GM_{sys}}{r^2}+\frac{c}{\gamma_0 t}\right)\frac{\mathbf{r}}{r}  
\\ \label{eq:RAR_aNb}
& \approx & 
a_N  - \frac{c}{\gamma_0 t}\frac{\mathbf{r}}{r} - \frac{r}{\gamma_0t^2} e_t\,,
\end{eqnarray}where $r_M \equiv  2GM_{sys}/c^2 \ll r$ is the Schwarzschild radius, which is neglected compared to the spatial position $r$. That is, an acceleration anomaly is obtained mainly in the spatial direction, about $|\mathbf{a} - a_N| \approx \gamma_0^{-1}c/t$ for $\mathbf{a} 
 \equiv a^r e_r$. However, the total acceleration also has a time-like component, that is, in the direction $e_t$. In particular, for a circular orbit with radius $r$, and taking into account the non-zero temporal contribution to the acceleration in the hyperconical universe with radius $ct$ \citep{Monjo2023}, the total centrifugal acceleration is \begin{eqnarray} \nonumber
 \frac{v^2}{c^2} {e_s} &=& -\left(ct e_t e^t + x^i e_i e^i\right) \frac{d^2{\hat s}}{c^2dt^2} 
\;\; \approx 
\\
&\approx & 
 \,ct\left( \frac{r}{\gamma_0c^2t^2}\right) \,e_t \; + \;  \left(\frac{GM_{sys}}{c^2r^2}+\frac{1}{\gamma_0 ct}\right)\frac{x^i x_i}{r} \,e_i\,,\hspace{3mm}\;\;\;\;\;   
\end{eqnarray}where $e_s$ is an effective space-like direction ($||e_s||^2 = e_se^s = -1$), while the absolute value of the velocity is given by\begin{eqnarray} \nonumber
 \frac{v^4}{c^4} &=&
  - \bigg|\bigg| \frac{v^2}{c^2}  {e_s} \bigg|\bigg|^2  \approx  \left(\frac{GM_{sys}}{rc^2}\right)^2 +\frac{2GM_{sys}}{\gamma_0 t c^3} \Longrightarrow 
  \\ \label{eq:limits}
  & \Longrightarrow & \begin{cases}
      v \approx \sqrt{\frac{GM_{sys}}{ r}}\;\;  \hspace{5mm} \mathrm{if}\; \frac{GM_{sys}}{r^2} >> \frac{2c}{\gamma_0 t} =:  a_{\gamma 0}
      \\ 
v \approx \sqrt[4]{\frac{2GM_{sys}c}{\gamma_0 t}}\;\;   \hspace{2mm}  \mathrm{if}\; \frac{GM_{sys}}{r^2} << \frac{2c}{\gamma_0 t} =  a_{\gamma 0}\,,
  \end{cases}
  \hspace{3mm}\;\;
\end{eqnarray}which satisfies two well-known limits of Newton's dynamics and Milgrom's (Equation~\ref{eq:limits} right), where $a_0$ is the Milogrom's acceleration parameter and $M_{sys} = M_{sys}(r)$ is the total mass within the central sphere of radius $r$. Finally, the velocity curve $v=v(r)$ can be reworded in terms of the Newtonian circular speed $v_N \equiv \sqrt{GM_{sys}(r)/r}$. Therefore, the predicted mass-discrepancy acceleration relation for rotation curves is \begin{eqnarray} \label{eq:RAR_aNc}
\left(\frac{v}{ v_N}\right)^2 \approx \sqrt{1 + \frac{1}{|a_N|} \frac{2c}{\gamma_0 t}} \; \Longrightarrow\;  \frac{a_C}{a_N} \approx \sqrt{1 + \frac{1}{|a_N|} \frac{2c}{\gamma_0 t}}\,, \hspace{1mm}\;
\end{eqnarray} where $a_C = v^2/r$ is the total radial acceleration and $a_N = GM_{sys}/r^2$ is the Newtonian acceleration. However, the absence of rotation in galaxy clusters leads to a radial acceleration similar to Equation~\ref{eq:RAR_aNb}. In any case, the projection factor $\gamma_0^{-1} = \gamma_{sys}^{-1}\cos \gamma_{sys}$ depends on the projective angle $\gamma_{sys}$, which can be estimated from the galaxy cluster approach (Equation~\ref{eq:model_clustb}) or from the general model (Equation~\ref{eq:model_generalb}), respectively, as follows: \begin{eqnarray} \nonumber
&   \sin^2 \gamma_{sys}(r) \approx 
\\ \label{eq:model_clust2b}
&   \;\; \;\;\approx \sin^2\gamma_{cen} - \left({\sin^2\gamma_{cen} -  \sin^2 \gamma_U} \right) \frac{2v_N^2(r)}{2v_N^2(r) + \varepsilon^2_H v_H^2(r)}\,, \;\; \hspace{5mm}
\\ \nonumber
&  \sin^2 \gamma_{sys}(r)  \approx  
\\ \label{eq:model_general2b}
&  \;\; \;\;\approx
\sin^2\gamma_U + \left({\sin^2\gamma_{cen} -  \sin^2 \gamma_U} \right) \bigg| \frac{2v_N^2(r)-\epsilon^2_H v_H^2(r)}{2v_N^2(r) + \epsilon^2_H v_H^2(r)}\bigg|\,,\;\;
\end{eqnarray}where $\gamma_{cen}$ can be fixed to $\gamma_{cen} = \mathrm{\pi}/2$ to test the 1-parameter ($\epsilon_H$) general model of Equation~\ref{eq:model_general2b}, while this study assumes that $\{\epsilon_H, \gamma_{cen}\}$ are free in our 2-parameter model for clusters (Equation~\ref{eq:model_general2b}). Finally, the empty projective angle is usually set as $\gamma_U = \mathrm{\pi}/3$ \citep{MCS2023}, which produces a projection factor of $\gamma_U^{-1}\cos\gamma_U \approx \frac{1}{2}$.


\subsection{Individual fitting}
\label{sec:individual_fitting}

\begin{figure*}
 \centering
	\includegraphics[scale=0.88]{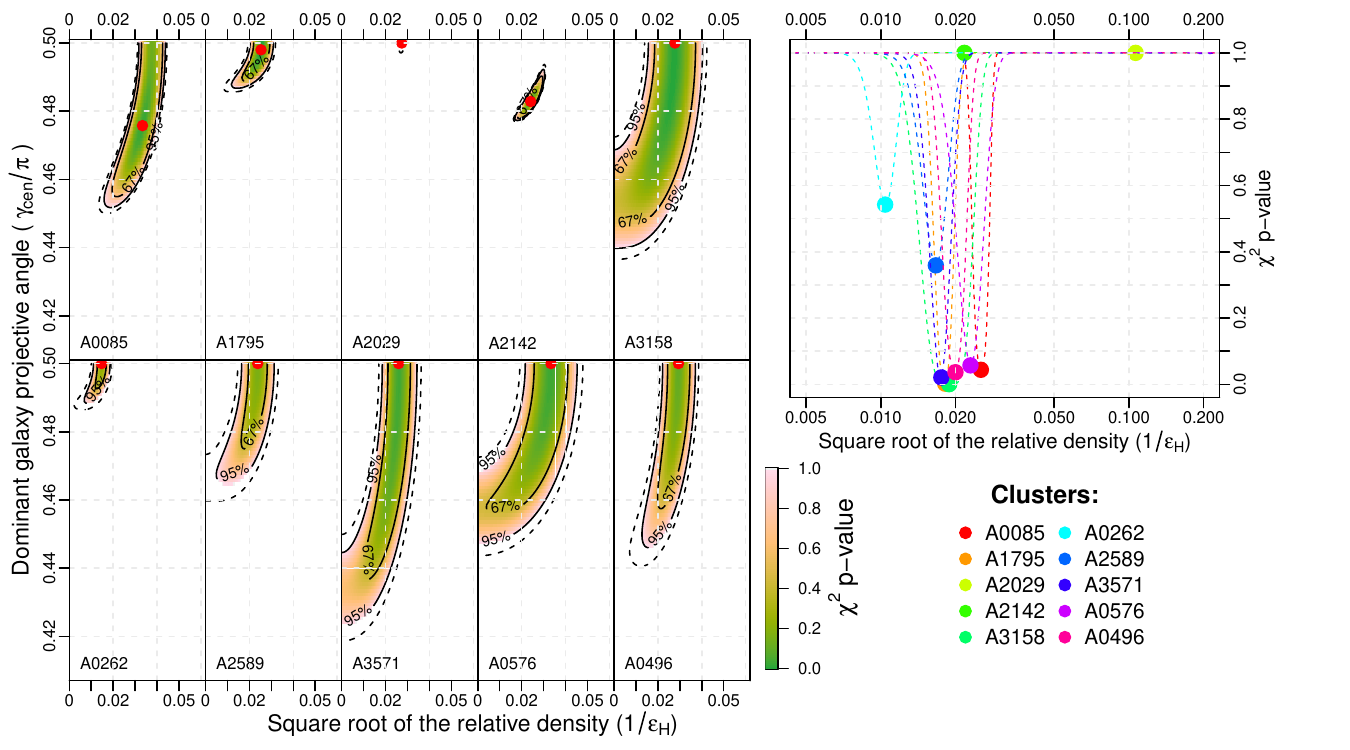}
    \caption{Observational constraints on the proposed models fitted to RAR data of the ten galaxy clusters considered \citep{Li2023}. \emph{Left}: Best values (red points) and uncertainty area (green shaded regions showing the $1\sigma$ confidence level) of the two parameters ($\varepsilon_H$ and $\gamma_{cen}$; see Table~\ref{table:table2}) used in fits to individual clusters according to the specific approach (Equation~\ref{eq:model_clust2b} is used in Equation~\ref{eq:RAR_aNb}). \emph{Right}: Best fit of the single free parameter ($\varepsilon_H$) used in the general model (Equation~\ref{eq:model_general2b} used in Equation~\ref{eq:RAR_aNb}), with fixed $\gamma_{cen} = 0.48\mathrm{\pi}$. In both cases (1 or 2 free parameters), the \textit{projective angle of the neighborhood} ($\gamma_U$) was fixed to $\gamma_U = \mathrm{\pi}/3$. Notice that cluster A2029 did not pass the $\chi^2$ test even with the 2-parameter model. However, it passed the test for $\gamma_U = 0.235\mathrm{\pi}$, corresponding to $\gamma_U^{-1}\cos{\gamma_U} \approx 1$.
}  \label{fig:FigC2}
\end{figure*}

Observed data on the RAR of 10 clusters ($0.0328 < z < 0.0899$) were collected from the study performed by \citet{Li2023}. Individually, fitting of Equation~\ref{eq:model_clust2b} for the anomaly between the total spatial acceleration and Newtonian acceleration (Equation~\ref{eq:RAR_aNb}) leads to a square root of the relative density of about $\varepsilon_H = 38^{+29}_{-11}$ (90\% confidence level, Figure~\ref{fig:FigC2}). All these results are obtained by fixing the constants $\gamma_U = \mathrm{\pi}/3$ and $\gamma_{cen} = \mathrm{\pi}/2$. The general model (Equation~\ref{eq:model_general2b}), with only one free parameter ($\varepsilon_H$), gave good results for eight of the ten clusters, showing difficulties in fitting the more available data from the A2029 and A2142 clusters (Table~\ref{table:table2}). If two parameters are considered ($\varepsilon_H$, $\gamma_{cen}$), the results considerably improve except for the A2029 cluster, which requires changing $\gamma_U^{-1}\cos\gamma_U \to 1$ to be compatibly fitted to the observations.

\begin{figure*}
\centering
\includegraphics[width=0.88\textwidth]{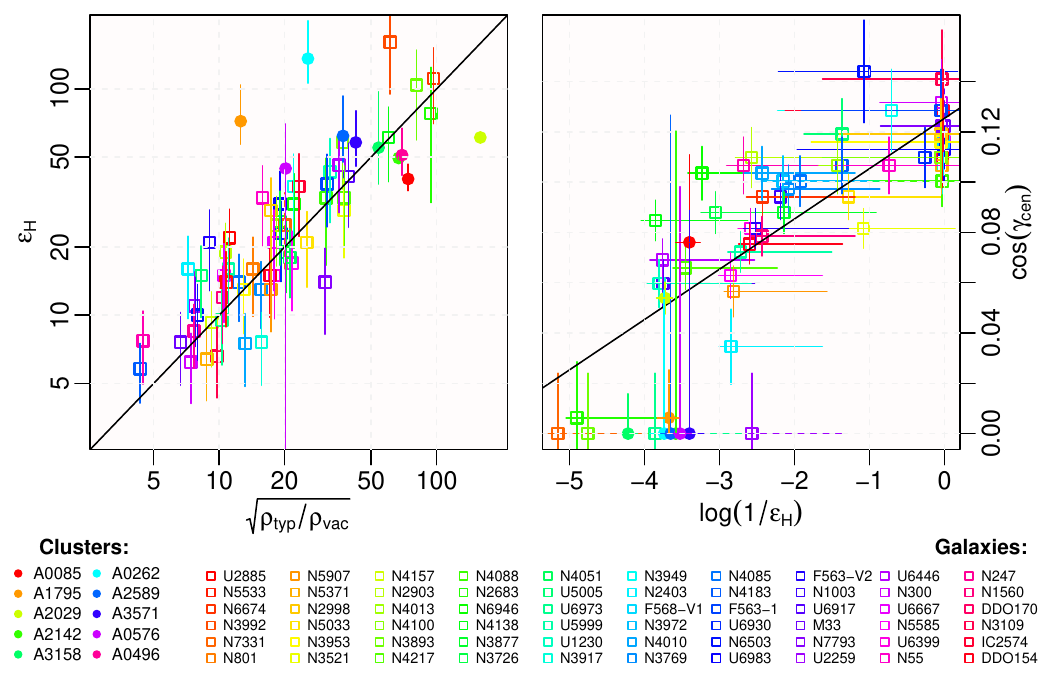}
    \caption{Model parameters fitted to the 60 high-quality galaxy rotation curves (squares), empirical relationships (black lines), and comparison with the fitting to the RAR data of 10 clusters (circles). \emph{Left}: Constraint on the 1-parameter HMG model for galaxies, with fixed $\gamma_{cen} = 0.48\mathrm{\pi}$ and free $\varepsilon_H$. This shows an empirical identity between the single parameter $\epsilon_H$ and the square root of the relative density $\rho_{typ}/\rho_{vac}$. In other words, in units of vacuum density $\rho_{vac} \equiv 3/(8\mathrm{\pi}Gt^2)$, an observed density $\rho_{typ} \equiv \rho(r_{typ})$ is defined at a typical equilibrium distance of $r_{typ} \sim 50-200\,$kpc, whose fit has been found ($R = 0.85$, $p$-value $<0.0001$) as four times the maximum radius ($r_{typ} \approx 4\,r_{max}$) for each galaxy rotation curve, and equal to the minimum radius ($r_{typ} \approx r_{min}$) for the data of each cluster. \emph{Right}: Constraint on the 2-parameter HMG model for galaxies, which shows a relationship between $\epsilon_H$ and $\gamma_{cen}$ such that $\cos(\gamma_{cen}) = \cos(0.460\mathrm{\pi}) - 0.020\ln\varepsilon_H$ for $1 \le \varepsilon_H < 400$, with a Pearson coefficient of $R = 0.80$ ($p$-value $<0.0001$; black regression line).}
    \label{fig:FigC3}
\end{figure*}

The same 2-parameter ($\varepsilon_H$, $\gamma_{cen}$) general model (Equation~\ref{eq:model_general2b} with $\gamma_U = \mathrm{\pi}/3$) was also applied to the 60 high-quality galaxy rotation curves, obtaining an acceptable result for all of them. The case of 1 parameter ($\varepsilon_H$ free when $\gamma_{cen} = 0.48\mathrm{\pi}$ is set) showed a slightly larger chi-square statistic and $p$-value, but these are also acceptable for all of them. Moreover, an empirical relationship is found between the single parameter $\epsilon_H$ and the square root of a relative density, which defines an identity $\rho(r_{typ})/\rho_{vac} \cong 1$ in units of vacuum density $\rho_{vac} \equiv 3/(8\mathrm{\pi}Gt^2)$ for an observed density $\rho(r_{typ})$ that is defined at a \textit{typical neighborhood} distance of approximately four times the maximum radius ($r_{typ} \approx 4\times r_{max}$, fitted at $R = 0.85$, $p$-value $<0.0001$, Figure~\ref{fig:FigC3} \emph{left}) for each galaxy rotation curve, and equal to the minimum radius ($r_{typ} \approx r_{min}$) for the data of each cluster. This \textit{typical distance} corresponds to $r_{typ} \approx 50-200\,$kpc.

Finally, when the 2-parameter HMG model is considered for galaxies, an additional relationship is found between $\epsilon_H$ and $\gamma_{cen}$:\begin{eqnarray}
    \cos \gamma_{cen} = \cos\left(0.4610^{+0.0013}_{-0.0014}\mathrm{\pi}\right) - (0.020 \pm 0.002)\ln \varepsilon_H\;\;
\end{eqnarray}for $1 \le \varepsilon_H < 400$, with a Pearson coefficient of $R = 0.80$ ($p$-value $<0.0001$, Figure~\ref{fig:FigC3} \emph{right}).

\begin{table}
    \setlength\extrarowheight{5pt}
    \caption{Individual fitting of Equation~\ref{eq:model_clust2} to each cluster according to the general model (Equation~\ref{eq:model_general2}) with one parameter ($\varepsilon_H$), and with the specific model for clusters (Equation~\ref{eq:model_clust2} with two parameters, namely $\varepsilon_H$ and $\gamma_{cen}$). The $p$-value is given for the lowest $\chi^2$.}
    \label{table:table2}
    \begin{tabular}{|l|cc|ccc|}
    \hline
    \multirow{2}{*}{Name (data)} & \multicolumn{2}{c|}{General model} & \multicolumn{3}{c|}{Specific model for clusters} \\ 
    & $\varepsilon_H$ & $\chi^2$ $p$-val & $\varepsilon_H$ & $\gamma_{cen}/\mathrm{\pi}$ & $\chi^2$ $p$-val \\ \hline \hline
    A0085 (17) & $40_{-5}^{+6}$ & 0.04 & $31_{-4}^{+5}$ & $0.474_{-0.009}^{+0.017}$ & $<0.01$  \\ 
    A1795 (4) & $55_{-9}^{+14}$ & $<0.01$ & $39_{-4}^{+6}$ & $0.499_{-0.006}^{+0.001}$ & $<0.01$  \\
    A2029 (32) & $52_{-20}^{+40}$ & $>0.95$ & $37_{-10}^{+15}$ & $0.500_{-0.002}^{+0}$ &  $>0.95$ \\
    A2142 (31) & $46_{-15}^{+15}$ & $>0.95$ & $41_{-2}^{+4}$ & $0.483_{-0.002}^{+0.003}$ & $<0.01$ \\
    A3158 (7) & $53_{-17}^{+39}$ & $<0.01$ & $37_{-8}^{+18}$ & $0.491_{-0.030}^{+0.009}$ & $<0.01$ \\
    A0262 (3) & $96_{-19}^{+32}$ & 0.54 & $67_{-6}^{+10}$ & $0.500_{-0.005}^{+0}$ & 0.57 \\
    A2589 (3) & $60_{-14}^{+37}$ & 0.36 & $41_{-4}^{+10}$ & $0.500_{-0.013}^{+0}$ & 0.47 \\
    A3571 (3) & $57_{-12}^{+22}$ & 0.02 & $39_{-6}^{+8}$ & $0.495_{-0.036}^{+0.005}$ & 0.14 \\
    A0576 (3) & $44_{-11}^{+22}$ & 0.06 & $31_{-6}^{+8}$ & $0.493_{-0.023}^{+0.007}$ & 0.13 \\
    A0496 (5) & $50_{-9}^{+16}$ & 0.04 & $33_{-4}^{+6}$ & $0.500_{-0.028}^{+0}$ & 0.14 \\ \hline
    \end{tabular}
\end{table}

\end{appendix}

\bsp
\label{lastpage}
\end{document}